\documentclass[useAMS,usenatbib]{mn2e}
 \usepackage{graphicx,amsmath,float,amssymb,bm,xcolor}
 \usepackage{rotating,lscape,url}
\title[Type Ia supernova SN 2014J]{Optical and NIR observations of the nearby type Ia supernova SN 2014J}

\author[Srivastav et al.]{Shubham Srivastav$^1$\thanks{E-mail : ssrivastav@iiap.res.in}, J.~P.\ Ninan$^2$\thanks{E-mail : ninan@tifr.res.in}, 
B.\ Kumar$^1$, G.~C.\ Anupama$^1$, D.~K.\ Sahu$^1$, 
\newauthor 
D.~K.\ Ojha$^2$, T.~P.\ Prabhu$^1$ \\
$^1$Indian Institute of Astrophysics, II Block Koramangala, Bangalore 560 034, India \\
$^2$Department of Astronomy and Astrophysics, Tata Institute of Fundamental Research, Homi Bhabha Road, Colaba, Mumbai - 400 005, India}

\begin{document}

\maketitle

\begin{abstract}
Optical and NIR observations of the type Ia supernova SN 2014J in M82 are presented.
The observed light curves are found to be similar to normal SNe Ia, with a decline 
rate parameter $\Delta m_{15}(B) = 1.08 \pm 0.03$. The supernova reached $B$-band 
maximum on JD 2456690.14, at an apparent magnitude $m_B(max) = 11.94$.
The optical spectra show a red continuum with deep interstellar Na~{\sc i} absorption, but 
otherwise resemble those of normal SNe Ia. The Si~{\sc ii} $\lambda 6355$ feature 
indicates a velocity of $\sim 12\,000$ km s$^{-1}$ at $B$-band maximum, which places 
SN 2014J at the border of the Normal Velocity and High Velocity group of SNe Ia.
The velocity evolution of SN 2014J places it in the Low Velocity Gradient 
subclass, whereas the equivalent widths of Si~{\sc ii} features near $B$-band 
maximum place it at the border of the Core Normal and Broad Line subclasses of SNe 
Ia. An analytic model fit to the bolometric light curve indicates that a total of 
$\sim 1.3$ M$_{\odot}$ was ejected in the explosion, and the ejected $^{56}$Ni mass
$M_{Ni} \sim 0.6$ M$_{\odot}$. The low [Fe~{\sc iii}] $\lambda4701$ to [Fe~{\sc ii}] $\lambda5200$ ratio in the 
nebular spectra of SN 2014J hints towards clumpiness in the ejecta. 
Optical broadband, linear polarimetric observations of SN 2014J obtained on 
four epochs indicate an almost constant polarization ($P_{\rm R} \sim$2.7 per cent; 
$\theta$ $\sim$37$^\circ$), which suggests that the polarization signal is of interstellar origin.
\end{abstract}


\section{Introduction}

Type Ia supernovae (SNe Ia) are identified by the presence of a prominent Si~{\sc ii} absorption trough near 6100 \AA\ in their optical spectra \citep{Filippenko97}.
SNe Ia constitute a remarkably homogeneous subclass, which makes them excellent standardizable candles for measuring 
cosmic distances \citep{Riess98,Perlmutter99}. SNe Ia are thought to be thermonuclear 
disruptions of accreting carbon-oxygen white dwarfs (WDs) in a close binary 
system \citep{Hoyle60,Woosley86}. Plausible progenitor systems for SNe Ia include the single degenerate (SD) scenario
\citep{Whelan73} where the companion is a main sequence or giant star, and the double degenerate (DD) scenario 
\citep{Iben84,Webbink84} wherein the companion is also a WD (see \citealt{Maoz14} for a recent review).
The details of the explosion mechanism and the nature of the progenitor, however, still remain unclear.
The wealth of recent data on SNe Ia has revealed the inherent diversity within this class of objects. 
While the majority ($\sim 70 \%$) of SNe Ia fall under the ``normal'' category, sub-luminous 1991bg-like objects
make up $\sim 18 \%$, whereas over-luminous 1991T-like events and peculiar 2002cx-like events constitute the remaining
$12 \%$ \citep{Li11}. 

The uncertainty in the amount of extinction suffered by SNe Ia in their host galaxy environment is a significant
obstacle in using them effectively as cosmological probes. Highly reddened SNe Ia tend to show unusually low values
of $R_V \lesssim 2$ \citep{Nobili08,Phillips13}, eg. SN 1999cl \citep{Krisciunas06}, SN2002cv \citep{EliasRosa08},
SN 2003cg \citep{EliasRosa06} and 2006X \citep{Wang08}; as opposed to the average Galactic value of $R_V = 3.1$. 
The observational evidence suggests that SNe Ia generally exhibit very weak continuum 
polarization ($P \lesssim 0.3$ per cent) but a moderate level of line polarization up to a few per cent 
\citep[see,][and references therein]{Wang97p,Wang03p,Wang06p,Leonard05p,Chornock06p,Wang08p,Patat09p,Smith11p,Maund13p}.
The continuum and line polarization depend upon the geometry of explosion and distribution of matter around the 
SN, respectively \citep{Wang08p}. Therefore, study of polarization properties can provide information on
asymmetry and orientation and also shed light on the time-varying behaviour of the SN ejecta. 
Generally, a significant level of line and continuum polarization tends 
to be detected in these events at early times (pre or near maximum light), decreasing at later epochs
\citep[although there are exceptions, for example see][]{Zelaya13p}.

SN 2014J was serendipitously discovered by \citet{Fossey14} on 2014 January 21, in the 
nearby galaxy M82. The discovery prompted an international multi-wavelength 
observational effort, spanning radio to $\gamma$-ray observations. At a distance of 
only $\sim 3.5$ Mpc, SN 2014J is among the nearest SNe Ia ever observed and thus 
provides an excellent opportunity to study in detail the properties of the progenitor 
and the nature of absorbing  dust along the line of sight.

Pre-discovery observations were reported from the Katzman Automatic 
Imaging Telescope (KAIT) as part of the Lick Observatory Supernova Search (LOSS) by 
\citet{Zheng14}, who also constrained the time of first-light between January 14.54 
and 14.96 UT, with January 14.75 UT as the best estimate. Using high cadence 
pre-explosion data obtained by the Kilodegree Extremely Little Telescope (KELT), 
\citet{Goobar15} report an explosion date of January $14.54 \pm 0.3$ UT, 5 hours 
earlier than that reported by \citet{Zheng14}. 

Several detailed studies spanning a wide range of the electromagnetic spectrum have been published on SN 2014J. 
\citet{Goobar14} presented early optical and near-infrared (NIR) observations and noted the presence of high velocity
features of Si~{\sc ii} and Ca~{\sc ii} in the optical spectra and a best-fit value of $R_V = 1.4 \pm 0.15$ to explain 
the observations.
\citet{Amanullah14} presented ultraviolet (UV) to NIR photometry of SN 2014J up to $\sim 40$ days past $B$-band maximum
and inferred best-fit reddening parameter values of $E(B-V)_{true} \sim 1.3$ and $R_V \sim 1.4$ for a Galactic 
extinction law. The results of \citet{Amanullah14} were also found to be consistent with power-law extinction which was
invoked by \citet{Goobar08} to account for extinction caused by multiple scattering. 
Using multi-band photometry spanning UV to NIR, \citet{Foley14} reported similar reddening parameters of 
$E(B-V)_{true} \sim 1.2$ and $R_V \sim 1.4$. However, a two component model involving a typical dust component
($R_V \sim 2.5$) and a circumstellar scattering component provided a better fit to the data \citep{Foley14}.
Further, \citet{Gao15} used dust models comprising graphite and silicate components in order to model the color excess
toward SN 2014J and inferred $E(B-V)_{true} \sim 1.1$ and $R_V \sim 1.7$, consistent with the results of \citet{Amanullah14}
and \citet{Foley14}.
Using \emph{Swift}-UVOT data for SN 2014J, \citet{Brown15} concurred with the low 
value of $R_V$ derived by previous authors, but argued for an interstellar origin for the extinction rather 
than circumstellar. Although C~{\sc ii} is not detected in the early optical spectra \citep{Goobar14}, 
\citet{Marion15} reported detection of
the C~{\sc i} 1.0693 $\mu m$ feature in their NIR spectra. 
\citet{Jack15} presented a time series of high resolution spectra of SN 2014J and confirmed the presence of 
interstellar medium (ISM) features and diffuse interstellar bands (DIBs) in the spectra.
\citet{Kawabata14} and \citet{Patat15} present linear polarimetric observations of SN 2014J, which highlight the peculiar
properties of dust in M82 and argue that the major contribution to extinction comes from interstellar dust rather than 
circumstellar matter (CSM).
\citet{Crotts15} report detection of light echoes from SN 2014J, the most prominent ones originating at distances of 80 pc
and 330 pc from the SN.
An early detection of $\gamma$-ray lines from SN 2014J was reported by \citet{Diehl14}, while subsequent observations
of $\gamma$-ray emission due to $^{56}$Co were presented by \citet{Churazov14,Diehl15}.

Various lines of evidence seem to favor the DD progenitor scenario for SN 2014J.
Using pre-explosion archival \emph{HST} images, \citet{Kelly14} ruled out a bright red giant companion as the donor.
The non-detection of radio \citep{Perez-Torres14} and X-ray emission \citep{Margutti14} also rule out a majority of the
parameter space occupied by SD models.
\citet{Lundqvist15} failed to find traces of accreted material from a non-degenerate companion in nebular spectra,
strengthening the case for the DD scenario for SN 2014J.

We present in this paper the results of extensive monitoring of SN 2014J in optical 
and NIR bands undertaken at the Indian Astronomical Observatory (IAO) using the 
2-m Himalayan Chandra Telescope (HCT). Also presented are four epochs of broadband 
polarimetric observations obtained using the ARIES Imaging Polarimeter mounted at the 
1-m Sampurnanand Telescope (ST) at Manora Peak, Nainital.

\section{Observations and Reduction}

\subsection{Optical Photometry}

Photometric observations of SN 2014J commenced on 2014 January 24, $\sim 8$ days before $B$-band maximum, using the 
2-m Himalayan Chandra Telescope at the Indian Astronomical Observatory at Hanle, and continued till 2014 October 28.
The observations were made with the Himalayan Faint Object Spectrograph Camera (HFOSC). The SITe CCD available
with the HFOSC has an imaging area of 2K $\times$ 4K pixels, of which the central unvignetted 2K $\times$ 2K area was used
for imaging. The field of view in imaging mode is 10 $\times$ 10 arcmin$^2$, with an image scale of 0.296 
arcsec pixel$^{-1}$. The SN was imaged in the Bessell $UBVRI$ filters available with the HFOSC.
Landolt standard field PG0918+029 \citep{Landolt1992} was observed under photometric conditions on the nights of 
2014 January 30, February 24 and February 26 for photometric calibration of the supernova field.
 
\begin{figure}
\centering
\resizebox{0.85\hsize}{!}{\includegraphics{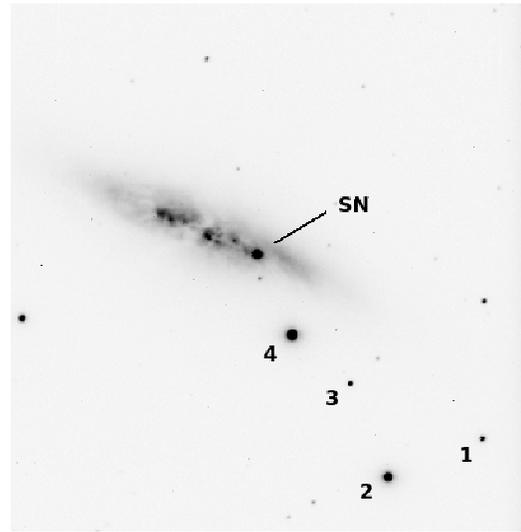}}
 \caption[]{\footnotesize Identification chart for SN 2014J. North is up and East is to the left. The field of view is
 $10' \, \times \, 10'$. Local standards are marked.}
 \label{fig:idchart}
\end{figure}

Data reduction was performed in the standard manner using various packages available 
with the Image Reduction and Analysis Facility (IRAF\footnote{IRAF is distributed by 
the National Optical Astronomy Observatories, which are operated by the Association 
of Universities for Research in Astronomy, Inc., under cooperative agreement with the 
National Science Foundation}). 
Aperture photometry was performed on the standard stars at two apertures - an aperture close to the FWHM of the stellar
profile, and another aperture of around 4 times the FWHM of the stellar profile. Aperture corrections were calculated as
the difference in magnitude for the two apertures. Average aperture corrections were calculated using
all the stars in the standard field.
Average extinction coefficients for the site \citep{Stalin08} were used in order to
account for atmospheric extinction, and average values of the color terms for the 
HFOSC system were used to arrive at the photometric solutions. The solutions thus 
obtained were used to calibrate local standards in the supernova field, 
observed on the same night as the standard fields. The local standards were thereafter
used to calibrate the supernova magnitudes. The field for SN 2014J is shown in 
Figure~\ref{fig:idchart}, and the magnitudes of the secondary standards are listed in 
Table~\ref{tab:secstd}. 

The magnitudes of the supernova and the secondary standards were evaluated using point-spread function (PSF)
fitting, with a fitting radius close to the FWHM of the stellar profile. 
Nightly photometric zero points were estimated using the local secondary standards and the supernova magnitudes were
evaluated differentially.

Since SN 2014J exploded in the interior of the bright starburst galaxy M82, there is 
significant contamination by the host galaxy in the late phase optical photometry. In 
order to subtract the host galaxy contribution, we used pre-explosion Sloan Digital 
Sky Survey (SDSS) $ugriz$ images of M82 as templates. In order to transform the images
from $ugriz$ to $BVRI$ filter system, we used the transformations prescribed by 
Lupton (2005)\footnote{\url{https://www.sdss3.org/dr8/algorithms/sdssUBVRITransform.php#Lupton2005}}.
The transformed SDSS images were then used as templates for host galaxy subtraction. \\
In order to check the robustness of this method, we applied the template subtraction to images obtained
near the epoch of maximum brightness, where the host galaxy contamination is expected to be minimal.
The magnitudes obtained using template subtraction were seen to show excellent agreement with those obtained using
PSF photometry near epochs of maximum brightness.
The magnitudes calculated using the above two methods indicate host galaxy contribution in the estimated SN
magnitudes beyond $\sim 60$ days since $B$-band maximum. Hence, beyond this epoch, we use the template-subtracted magnitudes.

Table~\ref{tab:mag} summarizes the optical $UBVRI$ photometric observations and also
lists the estimated magnitudes of SN 2014J.

\begin{table*}
\centering
\caption{$UBVRI$ Magnitudes of secondary standards in the field of SN 2014J}
\label{tab:secstd}
\vspace{3mm} 
\begin{tabular}{c c c c c c}
 \hline \hline
ID & U & B & V & R & I \\ 
\hline
 1 & 13.755 $\pm$ 0.010 & 13.912 $\pm$ 0.003 & 13.407 $\pm$ 0.002 & 13.095 $\pm$ 0.008 & 12.769 $\pm$ 0.011 \\
 2 & 12.113 $\pm$ 0.006 & 11.553 $\pm$ 0.007 & 10.715 $\pm$ 0.005 & 10.264 $\pm$ 0.001 & 9.871 $\pm$ 0.009 \\
 3 & 14.182 $\pm$ 0.010 & 14.351 $\pm$ 0.006 & 13.790 $\pm$ 0.005 & 13.458 $\pm$ 0.008 & 13.118 $\pm$ 0.004 \\
 4 & 10.761 $\pm$ 0.007 & 10.654 $\pm$ 0.004 & 10.052 $\pm$ 0.005 & 9.739 $\pm$ 0.026 & 9.452 $\pm$ 0.007 \\ 
\hline
\end{tabular}

\end{table*}

\begin{table*}
\centering
\caption{Optical and NIR photometry of SN 2014J}
\label{tab:mag}
\resizebox{\linewidth}{!}{%
\begin{tabular}{c c c c c c c c c c c}
\hline \hline
Date & JD & Phase$^*$ & $U$ & $B$ & $V$ & $R$ & $I$ & $J$ & $H$ & $K_s$ \\
(yyyy/mm/dd) & (245 6000+) & (days) & & & & & & & & \\
\hline 
2014/01/24 & 682.43 & $-$7.71 & 12.75 $\pm$ 0.01 & 12.45 $\pm$ 0.02 & 11.12 $\pm$ 0.02 & 10.49 $\pm$ 0.02 & 10.09 $\pm$ 0.02 & 9.55 & 9.60 & 9.48\\
2014/01/25 & 683.49 & $-$6.65 & 12.62 $\pm$ 0.01 & 12.30 $\pm$ 0.02 & 10.98 $\pm$ 0.01 & 10.39 $\pm$ 0.01 & 9.99 $\pm$ 0.01 & & & \\
2014/01/26 & 684.49 & $-$5.65 & 12.51 $\pm$ 0.02 & 12.20 $\pm$ 0.01 & 10.87 $\pm$ 0.01 & 10.31 $\pm$ 0.01 & 9.95 $\pm$ 0.02 & & &\\
2014/01/27 & 685.50 & $-$4.64 & 12.46 $\pm$ 0.02 & 12.10 $\pm$ 0.01 & 10.78 $\pm$ 0.01 & 10.25 $\pm$ 0.01 & 9.86 $\pm$ 0.02 & & &\\
2014/01/28 & 686.24 & $-$3.90 & 12.41 $\pm$ 0.01 & 12.07 $\pm$ 0.02 & 10.73 $\pm$ 0.02 & 10.21 $\pm$ 0.01 & 9.85 $\pm$ 0.01 & & &\\
2014/01/29 & 687.36 & $-$2.78 & 12.36 $\pm$ 0.01 & 12.01 $\pm$ 0.02 & 10.65 $\pm$ 0.01 & 10.13 $\pm$ 0.02 & 9.82 $\pm$ 0.01 & 9.35 & 9.51 & 9.24 \\
2014/01/30 & 688.38 & $-$1.76 & 12.36 $\pm$ 0.02 & 11.96 $\pm$ 0.01 & 10.61 $\pm$ 0.02 & 10.11 $\pm$ 0.02 & 9.81 $\pm$ 0.02 & 9.35 & 9.54 & 9.33 \\
2014/01/31 & 689.37 & $-$0.77 & 12.34 $\pm$ 0.01 & 11.94 $\pm$ 0.01 & 10.56 $\pm$ 0.01 & 10.13 $\pm$ 0.01 & 9.83 $\pm$ 0.02 & 9.43 & 9.62 & 9.36 \\
2014/02/01 & 690.41 & +0.27   & 12.38 $\pm$ 0.01 & 11.94 $\pm$ 0.01 & 10.55 $\pm$ 0.01 & 10.10 $\pm$ 0.01 & 9.85 $\pm$ 0.01 & & &\\
2014/02/02 & 691.06 & +0.92   & 12.41 $\pm$ 0.01 & 11.95 $\pm$ 0.01 & 10.54 $\pm$ 0.02 & 10.11 $\pm$ 0.02 & 9.86 $\pm$ 0.02 & & &\\
2014/02/03 & 692.29 & +2.15   &                  &                  & 10.56 $\pm$ 0.01 & 10.14 $\pm$ 0.01 & 9.92 $\pm$ 0.02 & & &\\
2014/02/05 & 694.17 & +4.03   & 12.56 $\pm$ 0.01 & 12.03 $\pm$ 0.01 & 10.57 $\pm$ 0.01 & 10.15 $\pm$ 0.01 & 9.97 $\pm$ 0.01 & & &\\
2014/02/07 & 696.42 & +6.28   & 12.69 $\pm$ 0.01 & 12.14 $\pm$ 0.01 & 10.61 $\pm$ 0.01 & 10.21 $\pm$ 0.02 & 10.05 $\pm$ 0.02 & & &\\
2014/02/09 & 698.46 & +8.32   & 12.84 $\pm$ 0.01 & 12.28 $\pm$ 0.02 & 10.72 $\pm$ 0.02 & 10.34 $\pm$ 0.01 & 10.17 $\pm$ 0.01 & & &\\
2014/02/11 & 700.28 & +10.14  & 13.00 $\pm$ 0.01 & 12.43 $\pm$ 0.01 & 10.83 $\pm$ 0.01 & 10.47 $\pm$ 0.02 & 10.30 $\pm$ 0.01 & 10.83 & 10.08 & 9.74 \\
2014/02/12 & 701.34 & +11.20  & 13.11 $\pm$ 0.01 & 12.52 $\pm$ 0.01 & 10.89 $\pm$ 0.01 & 10.54 $\pm$ 0.01 & 10.34 $\pm$ 0.01 & 11.08 & 10.02 & 9.80\\
2014/02/16 & 705.34 & +15.20  & 13.58 $\pm$ 0.01 & 12.92 $\pm$ 0.02 & 11.13 $\pm$ 0.004 & 10.78 $\pm$ 0.02 & 10.41 $\pm$ 0.01 & & &\\
2014/02/17 & 706.15 & +16.01  & 13.68 $\pm$ 0.02 & 12.99 $\pm$ 0.01 & 11.18 $\pm$ 0.01 & 10.78 $\pm$ 0.01 & 10.39 $\pm$ 0.02 & & &\\
2014/02/18 & 707.09 & +16.95  & 13.82 $\pm$ 0.01 & 13.13 $\pm$ 0.01 & 11.24 $\pm$ 0.01 & 10.80 $\pm$ 0.01 & 10.38 $\pm$ 0.01 & & &\\
2014/02/18 & 707.49 & +17.35  &                  & 13.17 $\pm$ 0.01 & 11.26 $\pm$ 0.01 & 10.80 $\pm$ 0.02 & 10.39$\pm$ 0.01 & & &\\
2014/02/19 & 708.38 & +18.24  & 13.98 $\pm$ 0.01 & 13.26 $\pm$ 0.01 & 11.27 $\pm$ 0.01 & 10.82 $\pm$ 0.01 & 10.35$\pm$ 0.01 & & &\\
2014/02/20 & 709.31 & +19.17  & 14.11 $\pm$ 0.02 & 13.37 $\pm$ 0.01 & 11.32 $\pm$ 0.01 & 10.81 $\pm$ 0.02 & 10.34 $\pm$ 0.01 & & &\\
2014/02/21 & 710.24 & +20.10  &                  & 13.45 $\pm$ 0.02 & 11.34 $\pm$ 0.02 & 10.80 $\pm$ 0.01 & 10.30 $\pm$ 0.01 & & &\\
2014/02/23 & 712.34 & +22.20  &                  & 13.67 $\pm$ 0.01 & 11.39 $\pm$ 0.01 & 10.80 $\pm$ 0.01 & 10.24 $\pm$ 0.01 & 11.24 & 9.79 & 9.66 \\
2014/02/24 & 713.32 & +23.18  & 14.58 $\pm$ 0.01 & 13.76 $\pm$ 0.01 & 11.43 $\pm$ 0.01 & 10.79 $\pm$ 0.01 & 10.22 $\pm$ 0.01 & & &\\ 
2014/02/25 & 714.32 & +24.18  & 14.69 $\pm$ 0.01 & 13.84 $\pm$ 0.01 & 11.47 $\pm$ 0.01 & 10.79 $\pm$ 0.02 & 10.20 $\pm$ 0.01 & 11.19 & 9.70 & 9.66\\
2014/02/26 & 715.32 & +25.18  & 14.79 $\pm$ 0.01 & 13.93 $\pm$ 0.01 & 11.51 $\pm$ 0.02 & 10.82 $\pm$ 0.01 & 10.19 $\pm$ 0.01 & & &\\
2014/02/27 & 716.30 & +26.16  &                  & 14.05 $\pm$ 0.01 & 11.55 $\pm$ 0.02 & 10.84 $\pm$ 0.01 & 10.18 $\pm$ 0.01 & 10.93 & 9.66 & 9.56 \\
2014/02/28 & 717.27 & +27.13  &                  &                  & 11.60 $\pm$ 0.01 & 10.86 $\pm$ 0.01 & 10.18 $\pm$ 0.01 & & &\\
2014/03/01 & 718.39 & +28.25  & 15.02 $\pm$ 0.02 & 14.20 $\pm$ 0.01 & 11.65 $\pm$ 0.01 & 10.91 $\pm$ 0.02 & 10.17 $\pm$ 0.01 & & &\\
2014/03/02 & 719.37 & +29.23  &                  &                  &                  &                  &                  & 10.74 & 9.69 & 9.61 \\
2014/03/03 & 720.31 & +30.17  & 15.16 $\pm$ 0.01 & 14.34 $\pm$ 0.01 & 11.73 $\pm$ 0.01 & 10.96 $\pm$ 0.01 & 10.20 $\pm$ 0.02 & 10.66 & 9.70 & 9.68\\
2014/03/04 & 721.33 & +31.19  & 15.22 $\pm$ 0.01 & 14.40 $\pm$ 0.01 & 11.80 $\pm$ 0.01 & 11.00 $\pm$ 0.01 & 10.20 $\pm$ 0.01 & 10.66 & 9.69 & 9.73 \\
2014/03/05 & 722.28 & +32.14  & 15.31 $\pm$ 0.02 & 14.47 $\pm$ 0.02 & 11.87 $\pm$ 0.01 & 11.05 $\pm$ 0.01 & 10.21 $\pm$ 0.03 & 10.59 & 9.78 & 9.75\\ 
2014/03/06 & 723.30 & +33.16  & 15.34 $\pm$ 0.02 & 14.52 $\pm$ 0.01 & 11.93 $\pm$ 0.01 & 11.13 $\pm$ 0.01 & 10.29 $\pm$ 0.02 & 10.65 & 9.79 & 9.80\\
2014/03/09 & 726.32 & +36.18  & 15.49 $\pm$ 0.01 & 14.68 $\pm$ 0.01 & 12.11 $\pm$ 0.01 & 11.34 $\pm$ 0.01 & 10.54 $\pm$ 0.02 & & &\\
2014/03/12 & 729.29 & +39.15  & 15.60 $\pm$ 0.01 & 14.81 $\pm$ 0.01 & 12.29 $\pm$ 0.01 & 11.53 $\pm$ 0.01 & 10.78 $\pm$ 0.02 & & &\\
2014/03/13 & 730.25 & +40.11  & 15.63 $\pm$ 0.02 & 14.83 $\pm$ 0.01 & 12.30 $\pm$ 0.01 & 11.60 $\pm$ 0.02 & 10.81 $\pm$ 0.01 & & &\\
2014/03/14 & 731.23 & +41.09  &                  & 14.88 $\pm$ 0.02 & 12.34 $\pm$ 0.01 & 11.61 $\pm$ 0.02 & 10.86 $\pm$ 0.02 & 11.40 & 10.33 & 10.42\\
2014/03/16 & 733.38 & +43.24  & 15.71 $\pm$ 0.02 & 14.90 $\pm$ 0.01 & 12.42 $\pm$ 0.01 & 11.70 $\pm$ 0.01 & 10.96 $\pm$ 0.01 & & &\\
2014/03/18 & 735.24 & +45.10  & 15.72 $\pm$ 0.02 & 14.94 $\pm$ 0.01 & 12.48 $\pm$ 0.01 & 11.80 $\pm$ 0.02 & 11.09 $\pm$ 0.02 & & &\\
2014/03/21 & 738.49 & +48.35  &                  & 14.97 $\pm$ 0.01 & 12.60 $\pm$ 0.01 & 11.96 $\pm$ 0.02 & 11.29 $\pm$ 0.02 & & &\\
2014/03/25 & 742.24 & +52.10  & 15.83 $\pm$ 0.01 & 15.03 $\pm$ 0.01 & 12.67 $\pm$ 0.01 & 12.04 $\pm$ 0.01 & 11.40 $\pm$ 0.02 & & &\\
2014/03/27 & 744.33 & +54.19  &                  & 15.05 $\pm$ 0.02 & 12.74 $\pm$ 0.01 & 12.13 $\pm$ 0.01 & 11.50 $\pm$ 0.02 & & &\\
2014/04/03 & 751.23 & +61.09  &                  & 15.16 $\pm$ 0.02 & 12.95 $\pm$ 0.02 & 12.37 $\pm$ 0.01 & 11.85 $\pm$ 0.03 & 12.89 & 11.30 & 11.28 \\
2014/04/11 & 759.18 & +69.04  & 16.12 $\pm$ 0.03 & 15.26 $\pm$ 0.02 & 13.21 $\pm$ 0.01 & 12.63 $\pm$ 0.02 & 12.17 $\pm$ 0.01 & & &\\
2014/04/17 & 765.37 & +75.23  &                  & 15.31 $\pm$ 0.01 & 13.33 $\pm$ 0.01 & 12.81 $\pm$ 0.01 & 12.30 $\pm$ 0.02 & & &\\
2014/04/27 & 775.34 & +85.20  &                  & 15.46 $\pm$ 0.01 & 13.60 $\pm$ 0.01 & 13.14 $\pm$ 0.02 & 12.78 $\pm$ 0.02 & & &\\
2014/04/30 & 778.16 & +88.02  &                  & 15.50 $\pm$ 0.01 & 13.68 $\pm$ 0.01 & 13.18 $\pm$ 0.01 & 12.86 $\pm$ 0.02 & 15.06 & 12.75 & 12.40\\
2014/05/03 & 781.31 & +91.17  & 16.62 $\pm$ 0.02 & 15.51 $\pm$ 0.02 & 13.75 $\pm$ 0.01 & 13.29 $\pm$ 0.01 & 12.98 $\pm$ 0.03 & & &\\
2014/05/07 & 785.25 & +95.11  &                  & 15.62 $\pm$ 0.01 & 13.88 $\pm$ 0.02 & 13.47 $\pm$ 0.03 & 13.15 $\pm$ 0.03 & & &\\
2014/05/27 & 805.26 & +115.12 & 17.20 $\pm$ 0.02 & 15.87 $\pm$ 0.05 & 14.39 $\pm$ 0.01 & 14.03 $\pm$ 0.02 & 13.80 $\pm$ 0.03 & & &\\
2014/06/04 & 813.24 & +123.10 &                  & 16.03 $\pm$ 0.02 & 14.57 $\pm$ 0.04 & 14.24 $\pm$ 0.02 & 14.00 $\pm$ 0.02 & & &\\
2014/06/16 & 825.13 & +134.99 &                  & 16.22 $\pm$ 0.02 & 14.88 $\pm$ 0.04 & 14.55 $\pm$ 0.04 & 14.23 $\pm$ 0.04 & & &\\
2014/07/05 & 844.17 & +154.03 &                  & 16.53 $\pm$ 0.02 & 15.29 $\pm$ 0.02 & 14.98 $\pm$ 0.01 & 14.60 $\pm$ 0.03 & & &\\
2014/07/21 & 860.17 & +170.03 &                  & 16.75 $\pm$ 0.02 & 15.57 $\pm$ 0.02 & 15.35 $\pm$ 0.02 & 14.91 $\pm$ 0.05 & & &\\
2014/09/27 & 928.43 & +238.29 &                  & 17.67 $\pm$ 0.02 & 16.59 $\pm$ 0.04 & 16.27 $\pm$ 0.04 & 15.71 $\pm$ 0.02 & & &\\
2014/10/28 & 959.46 & +269.32 &                  & 18.08 $\pm$ 0.04 & 16.99 $\pm$ 0.03 & 17.01 $\pm$ 0.03 & 16.08 $\pm$ 0.03 & & &\\
\hline
\multicolumn{3}{l}{$^*$\footnotesize{time since $B$-band max (JD 2456690.14)}}
   \end{tabular}}
   
\end{table*}  

\subsection{NIR Photometry}

NIR photometric observations of SN 2014J were carried out in \textit{J, H} and 
\textit{K$_S$} bands using the TIFR Near Infrared Spectrometer and Imager (TIRSPEC) 
mounted on HCT. TIRSPEC's detector is $1024 \times 1024$ Hawaii-1 PACE\footnote{
HgCdTe Astronomy Wide Area Infrared Imager -1, Producible Alternative to CdTe for 
Epitaxy} (HgCdTe) array. It has a pixel size of 18 $\mu m$ and a field of view (FoV) of
$\sim$ $5 \times 5$ arcmin$^{2}$. Broadband filters used for observations were $J$ ($\lambda_{center}$= 1.25 $\mu m$, 
$\Delta\lambda$= 0.16 $\mu m$), $H$ ($\lambda_{center}$= 1.635 $\mu m$, $\Delta\lambda$= 0.29 $\mu m$) and 
$K_S$ ($\lambda_{center}$= 2.145 $\mu m$, $\Delta\lambda$= 0.31 $\mu m$) 
(Mauna Kea Observatories Near-Infrared filter system).
Further details of TIRSPEC are available in \citet{Ninan14} and \citet{Ojha12}. 
A total of 17 nights of NIR photometric observations were obtained between 2014 January 24 and 
2014 April 30 (Table~\ref{tab:mag}).

Observations were carried out by pointing the telescope at 5 dither positions with 
15\arcsec\, step size. For generating a master sky frame, a separate sky region was 
also observed with same exposure and dither pattern immediately after each SN observation. A 
dark patch with no extended sources, 400\arcsec north of SN 2014J was chosen as the sky region. 
Data reduction was done using TIRSPEC photometry pipeline \citep{Ninan14}. The 
pipeline subtracts dark current for the equivalent exposure, and uses twilight flats 
of the same night for flat correction of data. Aperture photometry as well as PSF 
photometry were found to have significant contamination from the non-uniform
background of the host galaxy. Hence, 2MASS archive image of the galaxy was used as 
template and the galaxy background image was subtracted before aperture photometry. 
For aperture photometry, an aperture of 3 times FWHM was used. The uniform background 
sky was estimated from a ring outside the aperture radius with a width of 4.5\arcsec. 
2MASS magnitudes of secondary standards in the field were used for magnitude calibration and color correction.
Table~\ref{tab:mag} summarizes the $JHK_s$ magnitudes of SN 2014J.

\subsection{Optical Spectroscopy}

Optical spectroscopic observations of SN 2014J were made during 2014 January 24 to 2015 January 18, using grisms 
Gr7 (3500-7800 \AA) and Gr8 (5200-9250 \AA) available with the HFOSC. The details of the spectroscopic observations are 
summarized in Table~\ref{tab:speclog}.
Lamp spectra of FeAr and FeNe were used for wavelength calibration. The spectra were extracted in the standard manner.
Night sky emission lines $\lambda5577$, $\lambda6300$ and $\lambda6363$ were used to check the wavelength calibration 
and spectra were shifted wherever necessary. Spectra of spectrophotometric standards were observed in order to deduce 
instrumental response curves for flux calibration. For those nights where standard star observations were not available,
the response curves obtained during nearby nights were used. The flux calibrated spectra in grisms Gr7 and Gr8 were combined 
with appropriate scaling to give a single spectrum, which was brought to an absolute flux scale using the broadband 
$UBVRI$ photometric magnitudes. The spectra were corrected for redshift of the host galaxy. The telluric lines have not 
been removed from the spectra.

\begin{table*}
 \centering
 \caption{Log of spectroscopic observations of SN 2014J}
 \label{tab:speclog}
 \begin{tabular}{c c c c}
 \hline
 Date & JD & Phase$^*$ & Range \\
 (yyyy/mm/dd) & 245 6000+ & (days) & (\AA) \\
 \hline
 2014/01/24 & 682.43 & $-$7.71 & 3500-7800; 5200-9250 \\
 2014/01/25 & 683.49 & $-$6.65 & 3500-7800; 5200-9250 \\
 2014/01/26 & 684.49 & $-$5.65   & 3500-7800; 5200-9250 \\
 2014/01/27 & 685.50 & $-$4.64  & 3500-7800; 5200-9250 \\
 2014/01/28 & 686.24 & $-$3.90   & 3500-7800; 5200-9250 \\
 2014/01/29 & 687.36 & $-$2.78  & 3500-7800; 5200-9250; $YJHK$ \\
 2014/01/31 & 689.37 & $-$0.77  & 3500-7800; 5200-9250 \\
 2014/02/01 & 690.41 & +0.27   & 3500-7800; 5200-9250; $JK$ \\
 2014/02/02 & 691.06 & +0.92   & 3500-7800; 5200-9250 \\
 2014/02/07 & 696.42 & +6.28   & 3500-7800; 5200-9250 \\
 2014/02/11 & 700.30 & +10.16   & 3500-7800; 5200-9250; $YJHK$ \\
 2014/02/12 & 701.36 & +11.22   & 3500-7800; 5200-9250; $YJH$ \\
 2014/02/18 & 707.09 & +16.95   & 3500-7800; 5200-9250 \\
 2014/02/26 & 715.32 & +25.18   & 3500-7800; 5200-9250 \\
 2014/03/05 & 722.28 & +32.14   & 3500-7800; 5200-9250 \\
 2014/03/13 & 730.25 & +40.11   & 3500-7800; 5200-9250 \\
 2014/03/18 & 735.24 & +45.10   & 3500-7800; 5200-9250 \\
 2014/03/28 & 745.46 & +55.32   & 3500-7800; 5200-9250 \\
 2014/04/03 & 751.23 & +61.09   & 3500-7800; 5200-9250 \\
 2014/04/30 & 778.16 & +88.02   & 3500-7800; 5200-9250 \\
 2014/07/05 & 844.17 & +154.03   & 3500-7800; 5200-9250 \\
 2014/10/28 & 959.46 & +269.32   & 3500-7800; 5200-9250 \\
 2015/01/18 & 1041.24 & +351.09  & 3500-7800; 5200-9250 \\
 
 \hline
 \multicolumn{3}{l}{$^*$\footnotesize{time since $B$-band max (JD 2456690.14)}}
 \end{tabular}

\end{table*}

\subsection{NIR Spectroscopy}

NIR spectroscopic monitoring of SN 2014J from 1.02 $\mu m$ to 2.35 $\mu m$ was carried out using TIRSPEC.
Spectroscopic observations were carried out using 1.97\arcsec\, wide long slit with a resolution of R $\sim$ 1200.
Complete spectral range was covered in separate single order mode observations of $Y$, $J$, $H$ and $K$ orders.
Source spectrum was taken at least in two dither positions along the slit. Argon lamp spectrum for wavelength calibration
and Tungsten lamp for smooth continuum flat correction were taken immediately after each observation before moving any
of the optics elements like grism or slit. The nearby bright telluric standard star Alp Leo was also observed in exactly
the same configuration for telluric line correction immediately after SN 2014J observations. 
The log of NIR observations is listed in Table~\ref{tab:speclog}.

Spectroscopic data were reduced using TIRSPEC spectroscopy pipeline. Wavelength calibrated spectra were divided using the
telluric standard star spectra, and finally multiplied by a blackbody curve with a temperature of $11\,800$ K 
(telluric standard star temperature). The final spectra were flux calibrated using the magnitudes obtained from our 
photometric observations. $Y$ band spectra were separately calibrated by interpolating fluxes in $I$ and $J$ bands.

\subsection{Broadband Polarimetry}

\begin{table*}
\centering
\caption{Log of polarimetric observations and estimated polarization parameters of SN~2014J. }
\label{tab_log}
\begin{tabular}{ccc|cc|cc}
\hline \hline
\textsc{ut} Date &JD      & Phase$^*$ & Filter & $P \pm \sigma_{P}$  & $ \theta \pm \sigma_{\theta}$  \\
    (yyyy/mm/dd) &2456000+ &(days) &        & (per cent)          & ($^\circ)$           \\ \hline
2014/01/24 &682.37& $-7.77$ & $V$ & 3.77 $\pm$ 0.02&39.28 $\pm$ 1.43 \\
              &      &      & $R$ & 2.64 $\pm$ 0.06&34.85 $\pm$ 1.87 \\
              &      &      & $I$ & 1.59 $\pm$ 0.09&38.95 $\pm$ 1.73 \\
2014/02/05 &694.21&$+4.07$ & $R$ & 2.64 $\pm$ 0.05&38.24 $\pm$ 1.13 \\
2014/03/06 &723.18&$+33.04$ & $R$ & 2.67 $\pm$ 0.04&39.26 $\pm$ 1.45 \\
2014/03/30 &747.35&$+57.21$ & $R$ & 2.70 $\pm$ 0.04&33.72 $\pm$ 2.26 \\
\hline
\multicolumn{3}{l}{$^*$\footnotesize{time since $B$-band max (JD 2456690.14)}}
\end{tabular}  \\
\end{table*}

Broadband linear polarimetric observations of SN~2014J were made during four nights,
using the ARIES Imaging Polarimeter \citep*[AIMPOL,][]{Rautela04p} mounted at 
the Cassegrain focus of the 104-cm Sampurnanand telescope (ST) at Manora Peak, 
Nainital. On the first epoch, observations were obtained in the $VRI$ bands, but only in the $R$-band 
thereafter. The log of polarimetric observations is given in Table~\ref{tab_log}.

The polarimeter consists of a half-wave plate (HWP) modulator and a Wollaston prism 
beam-splitter. The Wollaston prism analyzer is placed at the backend of the telescope 
beam path in order to produce ordinary and extraordinary beams in slightly different 
directions separated by 28 pixels along the north-south direction on the sky plane. 
A focal reducer (85 mm, f/1.8) is placed between the Wollaston prism and the CCD 
camera. The AIMPOL camera is composed of 1024 pixels $\times$ 1024 pixels.
Each pixel corresponds to 1.73 arcsec, and the FoV is $\sim$8 arcmin 
in diameter on the sky. The FWHM of the stellar images varies from 2 to 3 pixels.
At each position of the half-wave plate (0$\degr$, 22.5$\degr$, 45$\degr$ and 67.5$\degr$),
multiple sets of frames were obtained. 

Fluxes of ordinary ($I_{o}$) and extra-ordinary ($I_{e}$) beams of the SN were 
extracted by standard aperture photometry after pre-processing using {\small IRAF}.
The detailed procedures used to estimate the polarization and polarization angles for 
the target objects can be found in \citet{Ramprakash98p} and \citet{Rautela04p}.

To correct our measurements for the zero-point polarization angle, polarized standard
stars \citep[BD~59389, HD~19820, HD~155197, HD~25443, HD~16056, HD~154445 and 
HD~25443; from][]{Schmidt92p} were observed on each night. The difference between the 
observed and standard values were applied to the SN. The instrumental 
polarization of AIMPOL on the 104 cm ST has been characterized for different projects 
and generally found to be $\sim$0.1 per cent \citep[e.g.,][and references therein]{Rautela04p,Pandey09p,Eswaraiah11p,Eswaraiah12p,Kumar14p}.
The SN polarization has been calibrated by applying this offset value. The estimated 
SN polarization parameters are listed in Table~\ref{tab_log}. 

\section{Optical and NIR Light Curves}\label{lc}

\begin{figure}
\centering
\resizebox{\hsize}{!}{\includegraphics{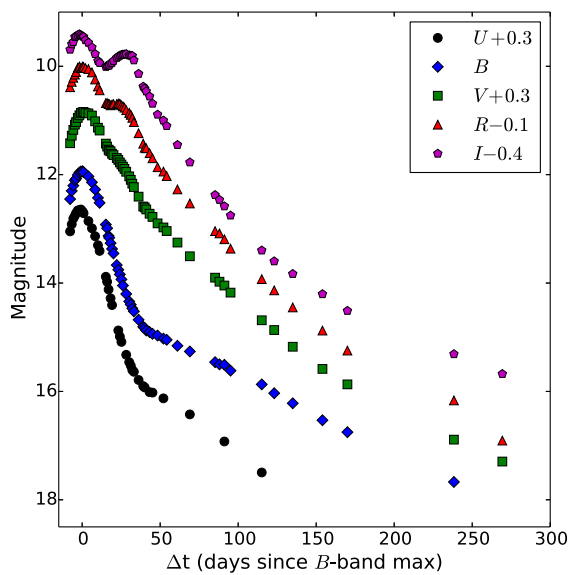}}
 \caption[]{\footnotesize Optical $UBVRI$ light curves of SN 2014J. The magnitudes past +60d are derived using template 
 subtraction. The typical errors on the magnitudes are smaller than the symbol size.}
 \label{fig:lc_ubvri}
\end{figure}

\begin{figure}
\centering
\resizebox{0.95\hsize}{!}{\includegraphics{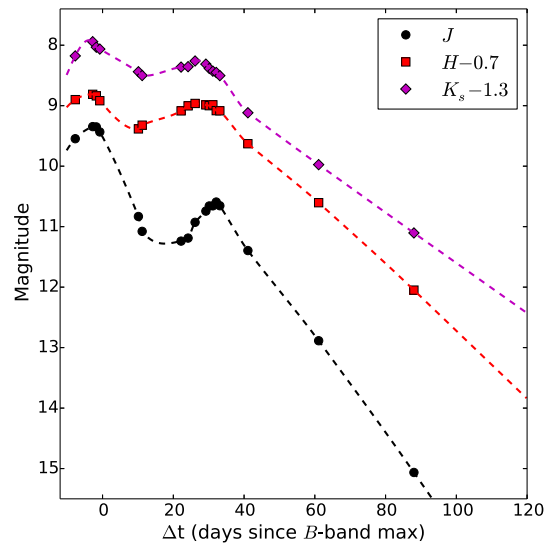}}
 \caption[]{\footnotesize NIR $JHK_s$ light curves of SN 2014J.}
 \label{fig:lc_jhk}
\end{figure}

SN 2014J was observed in the $UBVRI$ bands from $-8$d to $+269$d, and in the NIR 
$JHK_s$ bands from $-8$d to $+88$d since $B$-band maximum, which occurred on 
JD $2456690.14 \pm 0.5$ at an apparent magnitude of $11.94 \pm 0.02$.
The optical and NIR light curves of SN 2014J are shown in Figures~\ref{fig:lc_ubvri} and ~\ref{fig:lc_jhk},
respectively. The photometric properties of SN 2014J extracted from the light curves are summarized in 
Table~\ref{tab:phot_prop}.

SN 2014J suffers an unusually large amount of extinction in its host environment.
In addition to a single component dust model with low $R_V$, \citet{Foley14} had also invoked a
two component dust model which consists of a typical dust component along with a Large Magellanic Cloud-like dust
component in the circumstellar environment of SN 2014J in order to explain the observations.
However, polarization studies seem to rule out the presence of a CSM \citep{Kawabata14,Patat15}.
\citet{Brown15} also attribute most of the observed extinction to dust of interstellar rather than circumstellar origin.
Thus, in order to account for host extinction, we favor the single dust component model with
$E(B-V)_{true} = 1.24 \pm 0.10$ and $R_V = 1.44 \pm 0.06$ \citep{Foley14} for a standard CCM reddening
law \citep{Cardelli89}, as modified by \citet{ODonnell94}. 

The observed decline rate parameter, $\Delta m_{15}(B)_{obs} = 0.96 \pm 0.03$. However, the large amount of 
reddening effectively shifts the wavelength being sampled towards the red. The 
reddening-corrected decline rate parameter defined by \citet{Phillips99} is given by 
$$\Delta m_{15}(B)_{true} \simeq \Delta m_{15}(B)_{true} + 0.1E(B-V)_{true}$$ 
For $E(B-V)_{true} = 1.24$, we get $\Delta m_{15}(B)_{true} \simeq 1.08 \pm 0.03$, 
which is consistent with the estimates of \citet{Foley14,Ashall14,Tsvetkov14,Marion15}.

\begin{figure}
\centering
\resizebox{\hsize}{!}{\includegraphics{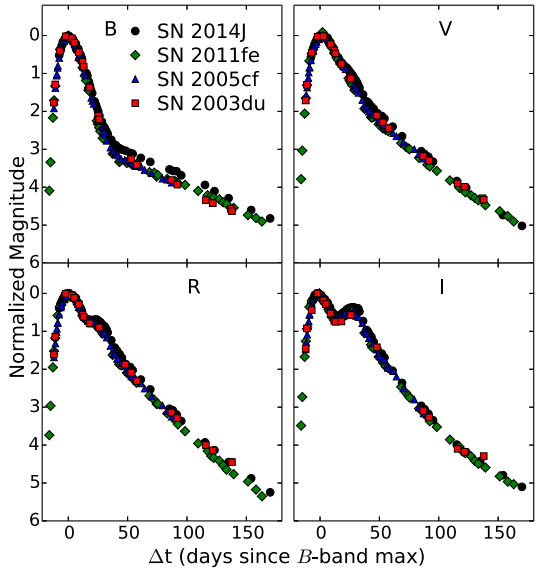}}
 \caption[]{\footnotesize Comparison of $BVRI$ light curves of SN 2014J with SN 2011fe, SN 2005cf and SN 2003du. The light
 curves of the SNe used for comparison have been shifted as described in the text.}
 \label{fig:lc_comp}
\end{figure}

Like in other normal SNe Ia, the light curves in the redder bands peak before $B$-band 
maximum. The optical $BVRI$ light curves of SN 2014J are compared with those of other normal 
SNe Ia like SN 2011fe \citep{Richmond12}, 2005cf \citep{Pastorello07} and 2003du 
\citep{Anupama05} in Figure~\ref{fig:lc_comp}. The SN magnitudes were normalized to 
the peak magnitudes in each filter and shifted in time to correspond to the epoch
of $B$-band maximum for SN 2014J. The $BVRI$ light curves of SN 2014J are remarkably 
similar to the other SNe Ia used for comparison.

\begin{table*}
 \centering
 \caption{Photometric properties of SN 2014J}
 \label{tab:phot_prop}
 \vspace{3mm}
 \begin{tabular}{c c c c c c c c}
 \hline
Band & JD (Max) & $m_{\lambda}^{max}$ & $M_{\lambda}^{max}$ & $\Delta m_{15}(\lambda)$ & Decline Rate (50-100d) & Decline Rate (100-250d)\\
      & 245 6000+&                    &                        &                       & mag (100d)$^{-1}$      & mag (100d)$^{-1}$ \\ 
 \hline
$U$  & 689.33 & 12.34 & $-$19.70 & 1.11 & 2.08 & - \\
$B$  & 690.14 & 11.93 & $-$19.09 & 1.08 & 1.38 & 1.43 \\
$V$  & 690.96 & 10.54 & $-$19.13 & 0.62 & 2.76 & 1.82 \\
$R$  & 689.60 & 10.10 & $-$18.84 & 0.64 & 3.24 & 1.93 \\
$I$  & 688.24 & 9.81  & $-$18.43 & 0.58 & 3.98 & 1.68 \\
$J$  & 687.81 & 9.34  & $-$18.68 & 1.83 & 8.04 & - \\
$H$  & 687.21 & 9.51  & $-$18.41 & 0.49 & 5.32 & - \\
$K_s$& 686.41 & 9.22  & $-$18.62 & 0.58 & 4.20 & - \\
 \hline
 \end{tabular}

\end{table*}

\section{Polarization Results}

\begin{figure}
\begin{centering}
\includegraphics[scale=0.4]{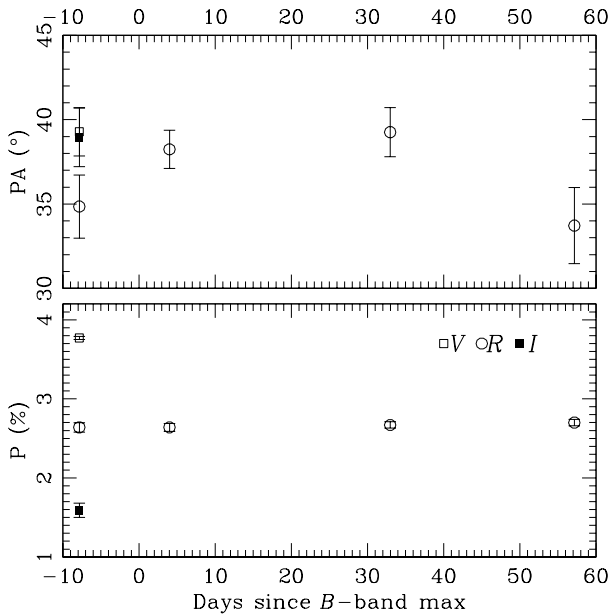}
\caption{Temporal evolution of the observed polarization parameters of SN~2014J. 
The upper and lower panels represent the polarization angle and degree of polarization,
respectively. Symbols used in both panels are same.}  
\label{phot_pol}
\end{centering}
\end{figure}

\begin{figure}
\begin{centering}
\includegraphics[scale=0.8]{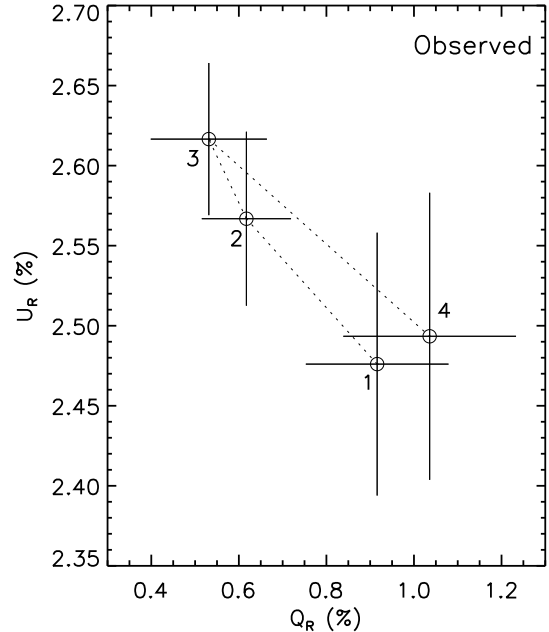}
\caption{Observed $Q$-$U$ diagram of SN~2014J in $R$-band.
Numbers labelled with 1--4 and connected with dotted lines, indicate the temporal order.}
\label{qu}
\end{centering}
\end{figure}

The observed polarization parameters of SN 2014J in different optical bands are shown in Figure~\ref{phot_pol}.
\citet[][hereafter K14]{Kawabata14} have used broadband optical and NIR polarimetric observations of 
SN 2014J obtained at five epochs to study the dust properties of M82 in the SN region. 
In general, our estimated results are consistent with K14 i.e. the polarization angle in $R$-band remains
almost constant (average value 36.52$\degr$ $\pm$ 3.46$\degr$) which indicates alignment with local spiral
structure of M82 \citep{Greaves00p}.
However, the average degree of polarization (in $R$-band) found in the present study is 
slightly higher than the estimated value of K14. Although our observations in $V$ and 
$I$ bands are restricted to a single epoch, the estimated polarization parameters in both 
studies are similar. 
In order to examine the simultaneous behaviour of the polarization and the polarization
angle of SN 2014J, the observed $Q$-$U$ Stokes parameters are plotted in Figure~\ref{qu}, where 
each number represents a data pair of the $Q$-$U$ vector in $R$-band. The limited data points, although indicative
of these vectors being aligned along a line (within observational uncertainties), are not sufficient to
draw conclusions on the geometry of the ejecta.

It is interesting to note that the photometric properties of SN 2014J show similarity with 
SN 2011fe, SN 2005cf and SN 2003du (see Sec~\ref{lc}). These SNe Ia showed relatively low host galaxy extinction
\citep[c.f.][]{Anupama05,Pastorello07,Patat13p}.
\citet{Smith11p} presented multi-epoch spectropolarimetric observations of SN 2011fe 
and inferred asymmetry in the photosphere with intrinsic continuum polarization between 0.2 -- 0.4 per cent. 
Additionally, they found a shift in polarization angle of 90$\degr$ during the follow-up 
period (up to $43$d after the explosion), implying a time-dependent large-scale asymmetry in 
the explosion. In a similar study of SN 2003du, \citet{Leonard05p} estimated an 
intrinsic polarization of 0.3 per cent and predicted clumps in the ejecta.

Based on the data set presented here, it is not possible to draw a firm conclusion about the SN ejecta and intrinsic
polarization properties of SN 2014J. However, the results show consistency with that of K14 and 
\citet{Patat15}.
As stated previously, there is no significant variation in the polarization parameters of this 
event. Nonetheless, as discussed by \citet{Maeda15}, it is possible that a combination of phenomena
(e.g. multiple scattering and ISM absorption) is responsible for such an abnormally high observed
degree of polarization for SN 2014J. 
The near constant polarization in $R$-band over nearly 70 days suggests that the polarization signal was of 
interstellar origin and the SN was likely minimally polarized.
Using high resolution spectra, \citet{Ritchey15} concluded a wide range of physical/environmental
conditions in the ISM of M82. 
The light echo properties of this event also indicate a complex structure for the interstellar dust in the host 
galaxy \citep{Crotts15}. 

\section{Bolometric Light Curve}

The quasi-bolometric light curve of SN 2014J was constructed using the broadband $UBVRIJHK_s$ photometry described in the
previous section. The broadband optical and NIR magnitudes were corrected for extinction assuming  $E(B-V)_{true} = 1.24$ 
for a Galactic extinction law with $R_V = 1.44$, which are the best-fit values obtained by \citet{Foley14}.
Following \citet{Foley14}, we use $E(B-V)_{MW} = 0.05$ to account for Milky Way reddening. \\
The extinction-corrected magnitudes were then converted to monochromatic fluxes using the zero points from 
\citet{Bessell98}. A spline curve was then fit through the monochromatic fluxes and the resulting curve was integrated 
under appropriate limits of wavelength as determined from the filter response curves, to yield the quasi-bolometric flux 
for the particular epoch. 
In order to estimate the UV contribution to the bolometric flux, we use the \emph{HST} spectra presented by \citet{Foley14}.
It may be noted that a significant portion of the HST UV spectra at the blue end had to be rejected owing to poor signal.
The UV contribution estimated here for SN 2014J might thus be treated as a lower limit.
We estimate that the UV contribution falls rapidly from $\sim 4 \%$ at $-5$d to $<1 \%$ beyond +20d for SN 2014J. 
The diminishing UV contribution is taken into account up to $\sim 30$ days since $B$-band maximum.
The peak contribution from UV to the bolometric flux was found to be $\sim 13 \%$ for SN 2011fe \citep{Pereira13}
and $\sim 19 \%$ for SN 2013dy \citep{Pan15}, which is higher than our estimate of $\sim 4 \%$ for SN 2014J 
using \emph{HST} spectra. 

$JHK_s$ photometry is not available beyond $+88$d, and $U$-band photometry is not available
beyond $+115$d. Thus, we use only the $UBVRI$ magnitudes to estimate the quasi-bolometric fluxes between +88 to +115d, 
and only $BVRI$ magnitudes thereafter.

\begin{figure}
\resizebox{\hsize}{!}{\includegraphics{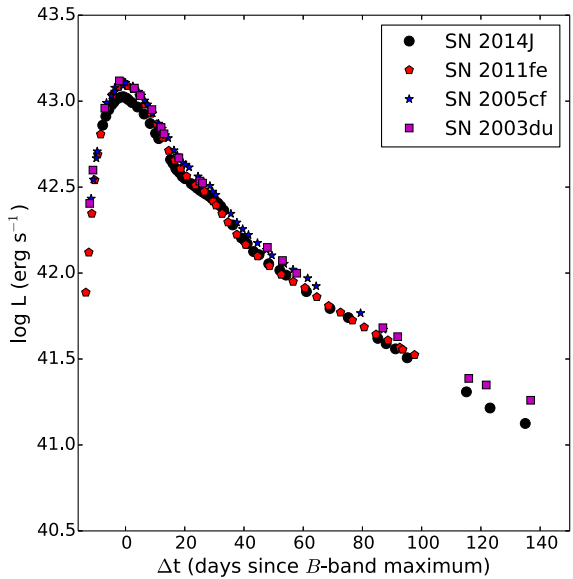}}
\caption[]{\footnotesize UVOIR bolometric light curve of SN 2014J plotted along with the bolometric light
curves of SN 2011fe, SN 2005cf and SN 2003du. The bolometric light curves were constructed as described in the text.}
 \label{fig:bol_comp}
\end{figure}

The UVOIR bolometric light curve of SN 2014J is shown in Figure~\ref{fig:bol_comp}, along with the
bolometric light curves of SN 2011fe, SN 2005cf and SN 2003du. The bolometric light curves of SNe 
2011fe \citep{Richmond12}, 2005cf \citep{Pastorello07} and 2003du \citep{Anupama05} were constructed using the 
published $UBVRI$ magnitudes. A 20 $\%$ constant contribution  was added in order to account for missing flux from UV 
and NIR passbands for SNe 2011fe, 2005cf and 2003du.

SN 2014J reached a peak UVOIR bolometric magnitude of $M_{bol}^{max} \approx -18.9 \pm 0.20$, which is similar to the peak
bolometric magnitudes of the normal SNe 2011fe, 2005cf and 2003du.
The date of explosion for SN 2014J was reported to be January 14.75 UT \citep{Zheng14} and 14.54 UT \citep{Goobar15}.
Assuming an explosion date of 14.65 UT, which is the average of the above estimates, the rise time to maximum for SN 2014J
is $\sim 18$ days, which is consistent with those of normal SNe Ia.

The bolometric light curves of SNe Ia are powered by the decay chain 
$^{56}$Ni $\rightarrow$ $^{56}$Co $\rightarrow$ $^{56}$Fe. A combination of three parameters - the mass of $^{56}$Ni
synthesized in the explosion ($M_{Ni}$), the total ejecta mass ($M_{ej}$) and the kinetic energy of the explosion ($E_k$)
determines the peak luminosity and width of the bolometric light curve \citep{Arnett82}. 

Using Arnett's rule \citep{Arnett82} to estimate the mass of $^{56}$Ni synthesized for a rise time $t_R \sim 18$ days and
a peak bolometric flux $L_{bol}^{max} \approx 1.06 \times 10^{43}$ erg s$^{-1}$, we obtain 
$M_{Ni} \approx 0.5$ M$_{\odot}$, assuming $\alpha = 1$.

In order to estimate the physical parameters of the explosion, i.e. $M_{Ni}$, $M_{ej}$ and $E_k$, we fit the observed
bolometric light curve with a simple analytical model proposed by \citet{Vinko04}. The assumptions in the model 
include spherical symmetry, homologous expansion of the ejecta and a constant opacity for $\gamma$-rays and positrons.
A core-shell density structure is assumed with a constant density core of fractional radius $x_0$, and an outer shell
with density decreasing outward as a power law with exponent $n$.

The free parameters in the model are ejected mass ($M_{ej}$), nickel mass ($M_{Ni}$), 
$\gamma$-ray opacity ($\kappa_{\gamma}$), positron opacity ($\kappa_{+}$), ejecta expansion velocity ($v_{exp}$) and the 
density power law exponent ($n$). 

Fitting this model to the post-maximum part of the bolometric light curve until $188$ days since explosion,
we obtain the following best-fit parameter values for a fixed fractional core radius ($x_0 = 0.15$): \\
$M_{ej} \sim 1.3$ M$_{\odot}$ (consistent with usual values of $0.9-1.4$ M$_{\odot}$ for SNe Ia), 
$M_{Ni} \sim 0.6$ M$_{\odot}$, $\kappa_{\gamma} \sim 0.03$ cm$^2$ g$^{-1}$ (consistent with
the usual value of 0.027 cm$^2$ g$^{-1}$ for grey atmospheres), $\kappa_{+} \sim 0.04$ cm$^2$ g$^{-1}$, 
$v_{exp} \sim 10\,000$ km s$^{-1}$ (consistent with photospheric velocity of $\sim 12\,000$ km s$^{-1}$ deduced from spectra)
and $n \sim 0.9$. The kinetic energy of the explosion, $E_k \sim 7 \times 10^{50}$ erg. \\

The $^{56}$Ni mass estimated above is consistent with the estimates of \citet{Churazov14}, \citet{Marion15} and
\citet{Telesco15} for SN 2014J. \citet{Childress15} estimate $M_{ej}$ and $M_{Ni}$ using [Co~{\sc iii}] 
$\lambda 5893$ emission feature in the nebular spectra of a sample of SNe Ia which includes SN 2014J. 
For SN 2014J, the estimated values are $M_{ej} \approx 1.44$ M$_{\odot}$ and $M_{Ni} \approx 0.84$ M$_{\odot}$.
\citet{Scalzo14a} estimated $M_{ej}$ and $M_{Ni}$ for a sample of 19 normal SNe Ia by modelling their bolometric light
curves and obtained a distribution of $0.9 < M_{ej} < 1.4$ M$_{\odot}$ and $M_{Ni}(avg) \sim 0.5$ M$_{\odot}$.

\begin{figure}
\resizebox{\hsize}{!}{\includegraphics{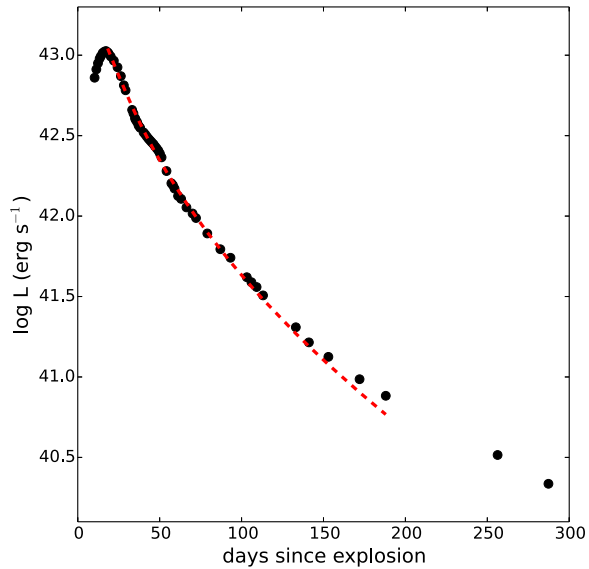}}
\caption[]{\footnotesize $UVOIR$ bolometric light curve of SN 2014J plotted along with the computed model
\citep{Vinko04} corresponding to the best-fit parameter values.}
 \label{fig:bol}
\end{figure}

\section{Spectroscopy Results}

Optical spectra of SN 2014J were obtained on 23 epochs between $-7.7$d to $+351.1$d relative to $B$-band maximum,
whereas NIR spectra were obtained on 4 epochs around $B$-band maximum.
The optical spectra of SN 2014J show a red continuum owing to the large amount of extinction suffered by the SN, but the 
spectral features are essentially those of a normal SN Ia.

\subsection{Spectral Evolution and SYN++ fits}

The photospheric phase spectra of SN 2014J show prominent features attributable to 
Si~{\sc ii}, S~{\sc ii}, Ca~{\sc ii}, Fe~{\sc ii} and Mg~{\sc ii} \citep[eg.][]{Goobar14}.
The similarity of the early spectra of SN 2014J to SN 2011fe
has been noted by \citet{Zheng14,Goobar14,Marion15}. However, unlike SN 2011fe, SN 2014J shows the presence of
high velocity features (HVFs) in pre-maximum spectra and lacks the C~{\sc ii} $\lambda 6580$ feature, which was detected
in SN 2005cf and SN 2011fe \citep{Zheng14,Goobar14}. 
The lack of detection of C~{\sc ii} in the optical spectra could be due to the rather late discovery of SN 2014J.
The earliest spectrum reported on January 22.3 UT \citep{Goobar14} corresponds to a phase of $\sim -9$d
relative to $B_{max}$. Detection of C~{\sc ii} becomes difficult at this phase, since the ionized carbon layer begins to
cool and recombine to form C~{\sc i}. This does not rule out the presence of carbon in the spectra however, as pointed out by
\citet{Marion15}, who report the detection of the C~{\sc i} 1.06 $\mu m$ feature in their early NIR spectra from 
$-7.4$d to $+3.2$d. \\

\begin{figure}
\centering
\resizebox{\hsize}{!}{\includegraphics{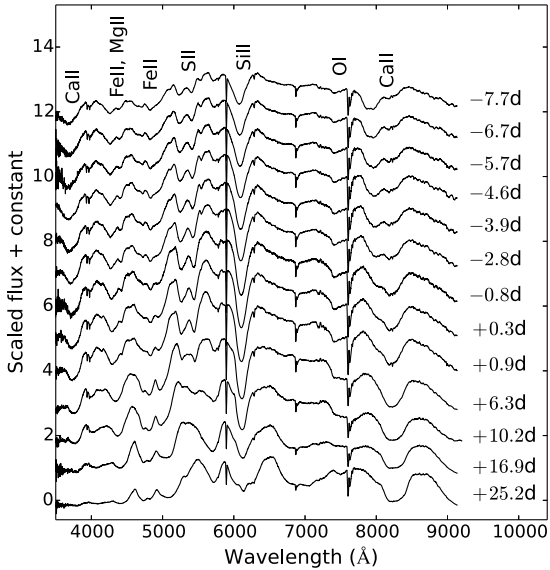}}
 \caption[]{\footnotesize Early spectral evolution of SN 2014J between $-7.7$d to $+25.2$d. Reddening correction has not 
 been applied to the spectra.}
 \label{fig:nearmax}
\end{figure}

\begin{figure}
\centering
\resizebox{\hsize}{!}{\includegraphics{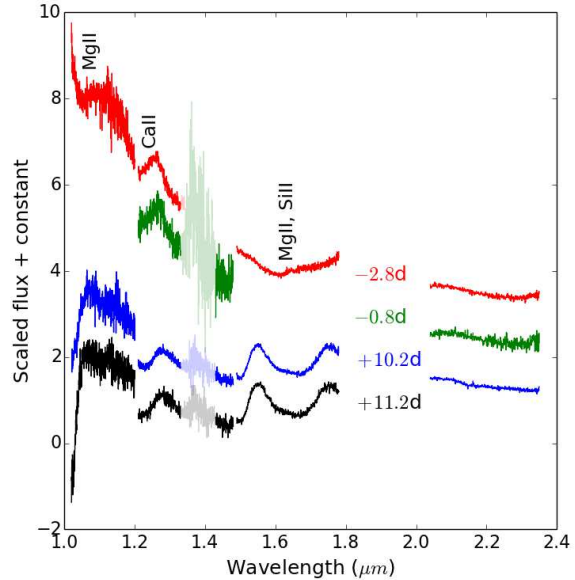}}
 \caption[]{\footnotesize Early NIR spectral evolution of SN 2014J between $-2.8$d to $+11.2$d.}
 \label{fig:nirspec}
\end{figure}

The early spectral evolution of SN 2014J in the optical ($-7.7$d to $+25.2$d) and NIR ($-2.8$d to $+11.2$d) is shown in
Figure~\ref{fig:nearmax} and Figure~\ref{fig:nirspec}, respectively.
The photospheric velocity of the ejecta, as measured from the blueshift of the Si~{\sc ii} $\lambda 6355$ absorption feature,
evolves from $\sim 13\,300$ km s$^{-1}$ at $-7.7$d to $\sim 12\,000$ km s$^{-1}$ near $B_{max}$ and further falls
to $\sim 11\,400$ km s$^{-1}$ at +17.0d, consistent with the measurements reported by \citet{Marion15}.
This places SN 2014J at the border of the normal velocity (NV) group and high velocity (HV) group of SNe Ia \citep{Wang09a}.
Our earliest optical spectrum ($-7.7$d) shows HV features of Ca~{\sc ii} detached at $\sim 24\,000$ km s$^{-1}$, but
this feature disappears rapidly as the epoch of $B$-band maximum approaches.
The earliest NIR spectrum presented here ($-2.8$d) exhibits a prominent Mg~{\sc ii} 1.09 $\mu m$ feature blueshifted at 
$\sim 13\,000$ km s$^{-1}$, whereas later NIR spectra are dominated by blended absorption due to
Mg~{\sc ii} 1.68 $\mu m$, Si~{\sc ii} 1.69 $\mu m$ features \citep{Marion15}.

\begin{figure}
\centering
\resizebox{\hsize}{!}{\includegraphics{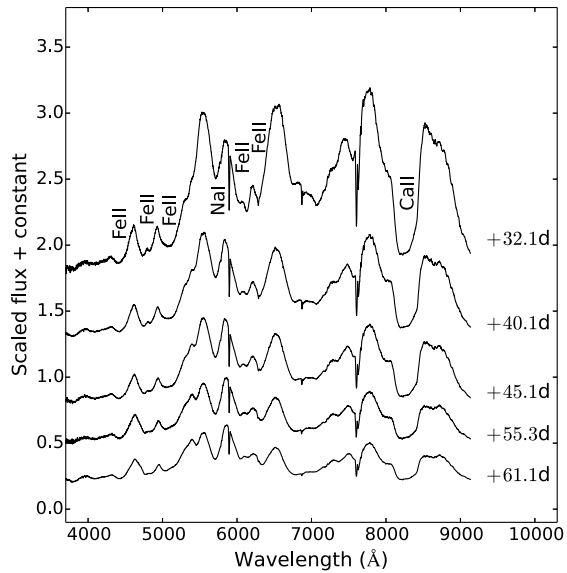}}
 \caption[]{\footnotesize Spectral evolution of SN 2014J between $+32.1$d to $+61.1$d. Reddening correction has not 
 been applied to the spectra.}
 \label{fig:post3060}
\end{figure}

The spectral evolution of SN 2014J between $\sim 1-2$ months since $B_{max}$ is presented in 
Figure~\ref{fig:post3060}. The Si~{\sc ii} $\lambda 6355$ feature has more or less
disappeared by $+32$d, and the spectra are progressively dominated by forbidden emission lines of [Fe~{\sc ii}] and
[Fe~{\sc iii}]. Other prominent features include the Na~{\sc i} D lines and the Ca~{\sc ii} NIR triplet, which
persists as a strong feature even at much later phases.

As the supernova enters the nebular phase, the ejecta becomes optically thin and transparent to optical photons,
thus exposing the Fe-rich core.
The nebular phase spectra of SN 2014J, from a phase of +88d to +351d are shown in Figures 15, 16 and 17. 
The +88d spectrum of SN 2014J is dominated by nebular emission features of [Fe~{\sc iii}] $\lambda 4701$, 
[Fe~{\sc ii}] $\lambda 5159$,$5261$ and Na~{\sc i}/[Co~{\sc iii}] features near 5900 \AA.
The [Fe~{\sc iii}] $\lambda 4701$ remains the most prominent feature in the +351d spectrum of SN 2014J. In addition, the 
[Fe~{\sc ii}] blend near 5200 \AA\ and the [Fe~{\sc ii}] $\lambda 7155$ and [Ni~{\sc ii}] $\lambda 7378$
features are also prominent.

We fit synthetic spectra using SYN++ to the early phase, dereddened spectra of SN 2014J. SYN++ is 
the rewrite of the SYNOW code in modern C++ \citep{Fisher00,Thomas11}.
The earliest spectrum obtained on $-7.7$d is fit with a photospheric velocity $v_{phot} = 15\,000$ km s$^{-1}$ and a 
blackbody temperature $T_{bb} = 13\,500$ K. The broad Ca~{\sc ii} NIR triplet is fit 
with two components -- 
a photospheric component and a HV component detached at $24\,000$ km s$^{-1}$.
The $-2.8$d spectrum is fit with the Ca~{\sc ii} HV component at a reduced velocity of $20\,000$ km s$^{-1}$. 
The HV feature is not seen in the spectra obtained beyond the epoch of $B$-band maximum.
The photospheric velocity decreases from $\sim 15\,000$ km s$^{-1}$ at $-7.7$d to $\sim 12\,000$ km s$^{-1}$ at $-2.8$d
and falls to $\sim 9\,000$ km s$^{-1}$ at +25.2d, whereas the blackbody temperature first increases slightly to 
$13\,800$ K at $-2.8$d and subsequently falls to $9\,000$ K at +25.2d.

\begin{figure*}
\hspace*{-5mm}
\centering
\resizebox{\hsize}{!}{\includegraphics{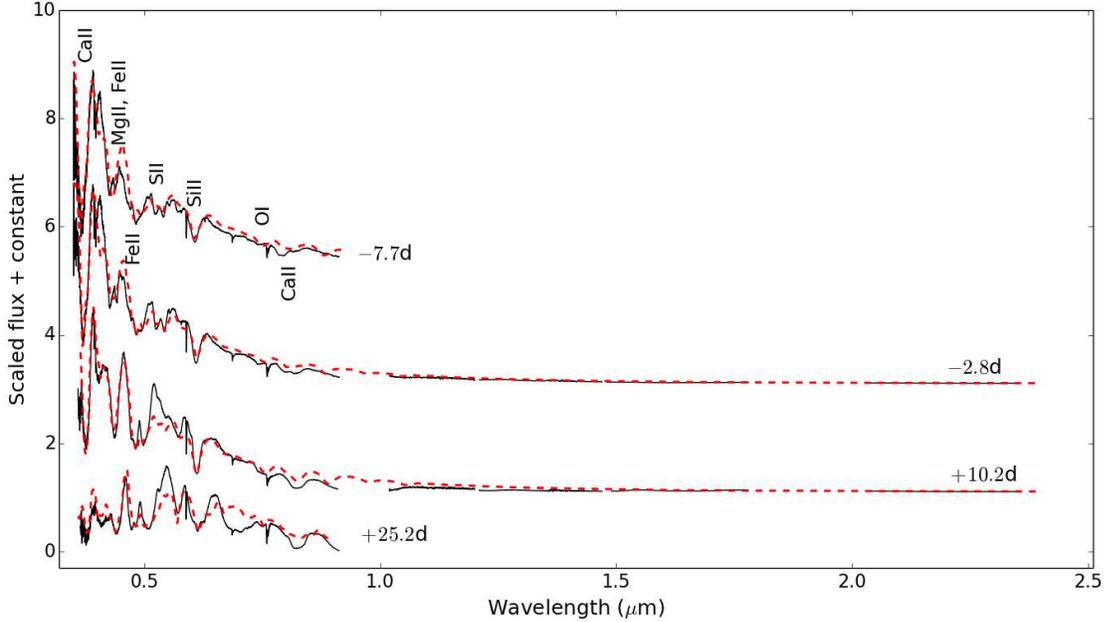}}
 \caption[]{\footnotesize Dereddened spectra of SN 2014J near the epoch of maximum light plotted along with synthetic
 spectra generated using SYN++ (dashed lines). The full optical/NIR fit is shown for the $-2.8$d and $+10.2$d spectra.}
 \label{fig:synfits}
\end{figure*}

Figure~\ref{fig:synfits} shows the SYN++ fits obtained for the $-7.7$, $-2.8$, $+10.3$ and $+25.2$d spectra of SN 2014J.

\subsection{Comparison with normal SNe Ia}

\begin{figure}
\centering
\resizebox{\hsize}{!}{\includegraphics{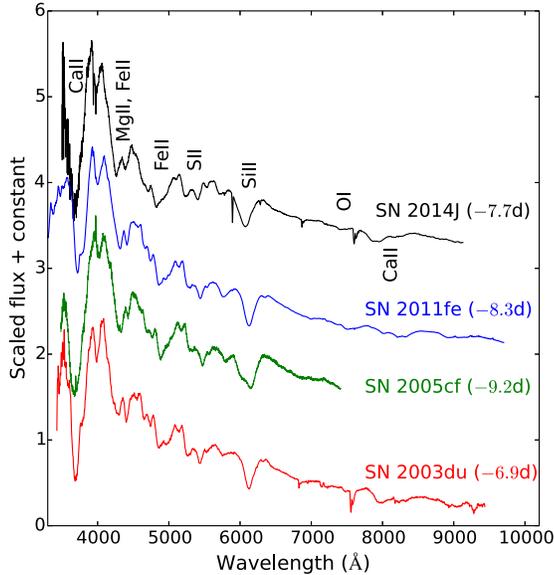}}
 \caption[]{\footnotesize Dereddened $-7.7$d spectrum of SN 2014J plotted with spectra of SN 2011fe, SN 2005cf and
 SN 2003du at similar epochs for comparison.}
 \label{fig:premaxcomp}
\end{figure}

Since SN 2014J has suffered significant reddening in its host environment, all the spectra of SN 2014J shown here
have been dereddened with $E(B-V) = 1.24$ and $R_V = 1.44$ \citep{Foley14} for comparison with SNe 2011fe, 2005cf and
2003du.
Figure~\ref{fig:premaxcomp} shows our earliest spectrum ($-7.7$d) of SN 2014J, plotted along with spectra of SN 2011fe 
\citep{Pereira13}, SN 2005cf \citep{Wang09b} and SN 2003du \citep{Anupama05} for comparison. The spectra are quite similar,
except for the fact that SN 2005cf shows a prominent HV component in the Si~{\sc ii} $\lambda 6355$ feature, unlike the other
SNe.

\begin{figure}
\centering
\resizebox{\hsize}{!}{\includegraphics{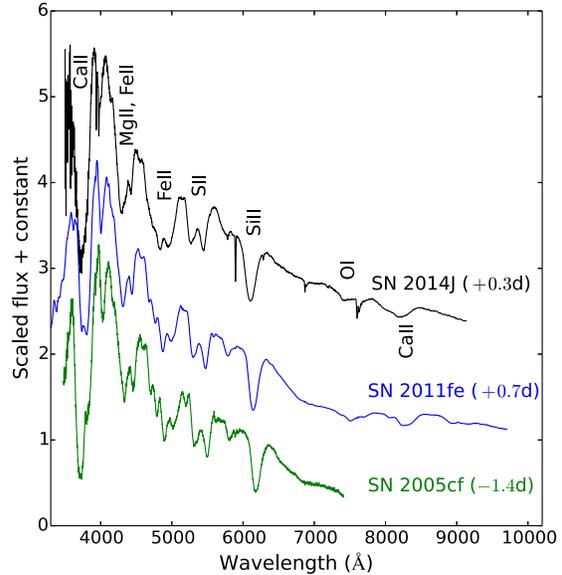}}
 \caption[]{\footnotesize  Dereddened spectrum of SN 2014J near epoch of B-band maximum plotted with spectra of SN 2011fe,
 SN 2005cf and SN 2003du at similar epochs for comparison.}
 \label{fig:maxcomp}
\end{figure}

Figure~\ref{fig:maxcomp} shows spectrum of SN 2014J near the epoch of $B$-band maximum, compared to SNe 2011fe and 2005cf.
Again, the spectra are very similar to each other but the features of SN 2014J show a higher blueshift, indicating higher
expansion velocities. 
The spectra of SN 2011fe and SN 2005cf used for comparison were downloaded from the WISeREP 
archive\footnote{\url{http://wiserep.weizmann.ac.il/}} \citep{Yaron12}.

\begin{figure}
\centering
\resizebox{\hsize}{!}{\includegraphics{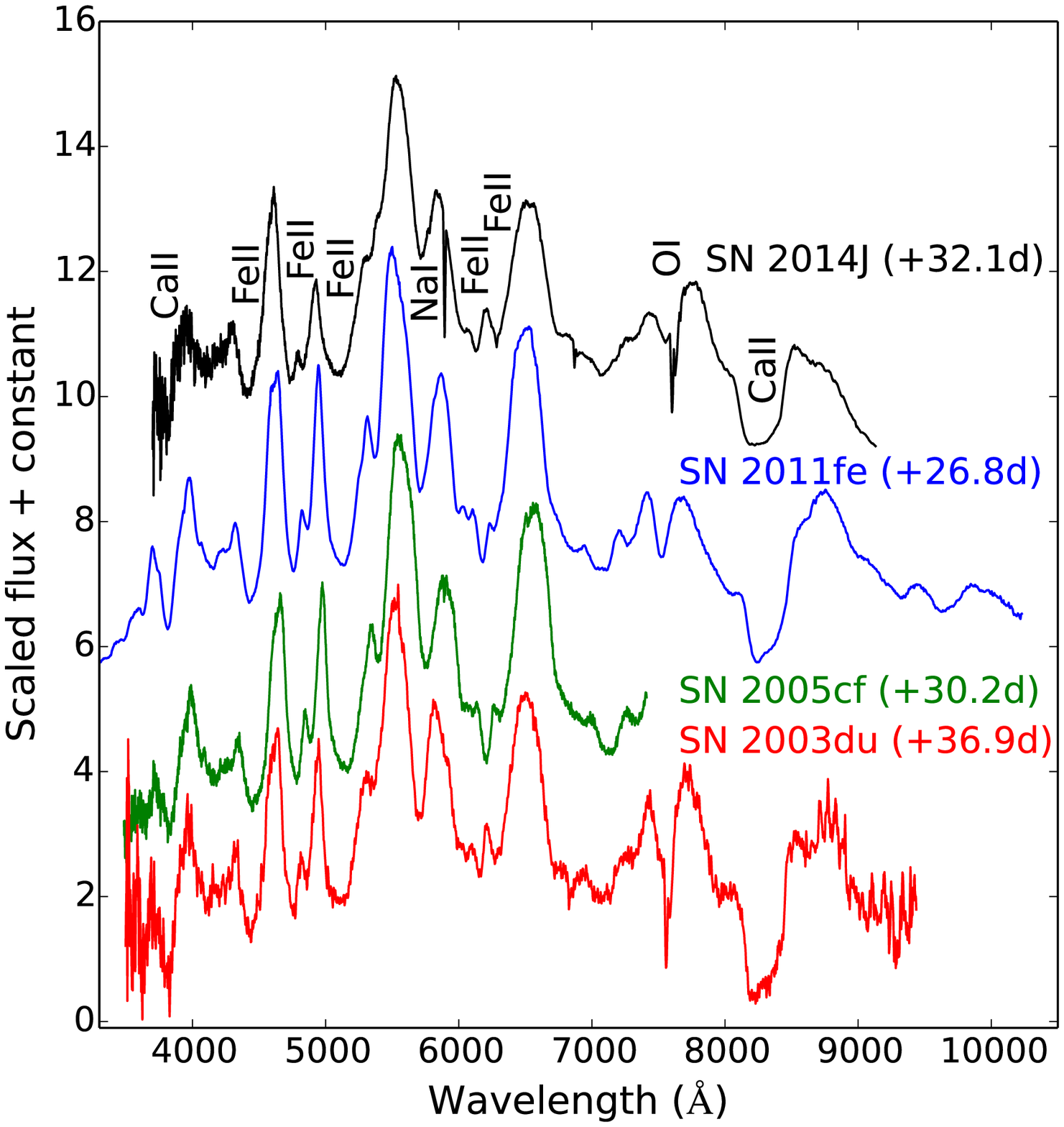}}
 \caption[]{\footnotesize Dereddened +32.1d spectrum of SN 2014J, plotted with spectra of SN 2011fe, SN 2005cf and
 SN 2003du at similar epochs.}
 \label{fig:p30comp}
\end{figure}

Figure~\ref{fig:p30comp} shows the +32.1d spectrum of SN 2014J, along with spectra of SN 2011fe \citep{Mazzali14},
SN 2005cf \citep{Wang09b} and SN 2003du \citep{Anupama05} at similar epochs.

\begin{figure}
\centering
\resizebox{\hsize}{!}{\includegraphics{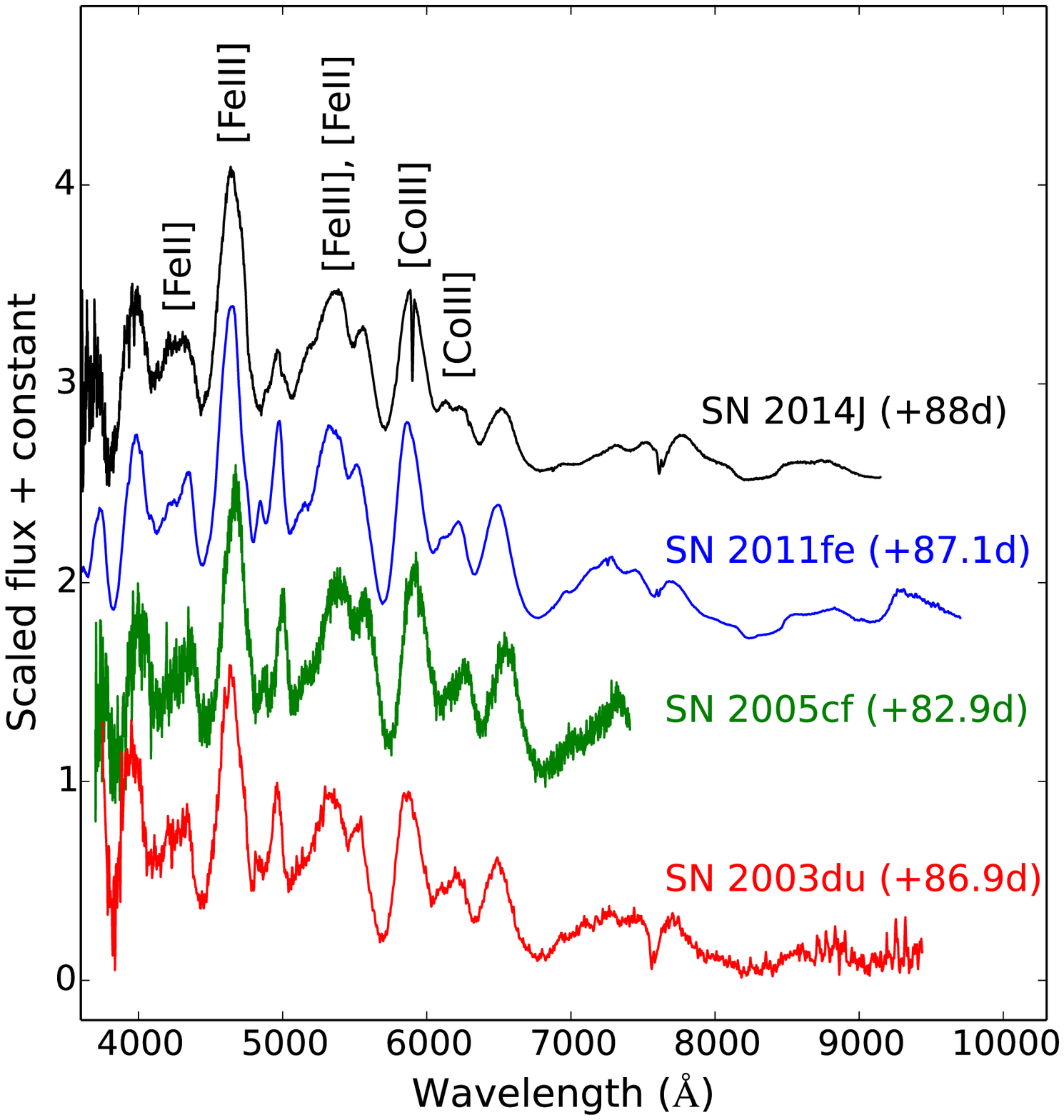}}
 \caption[]{\footnotesize Dereddened +88d spectrum of SN 2014J, plotted along with spectra of SNe 2011fe, 2005cf and 2003du
 at similar epochs.}
 \label{fig:p90comp}
\end{figure}

\begin{figure}
\centering
\resizebox{\hsize}{!}{\includegraphics{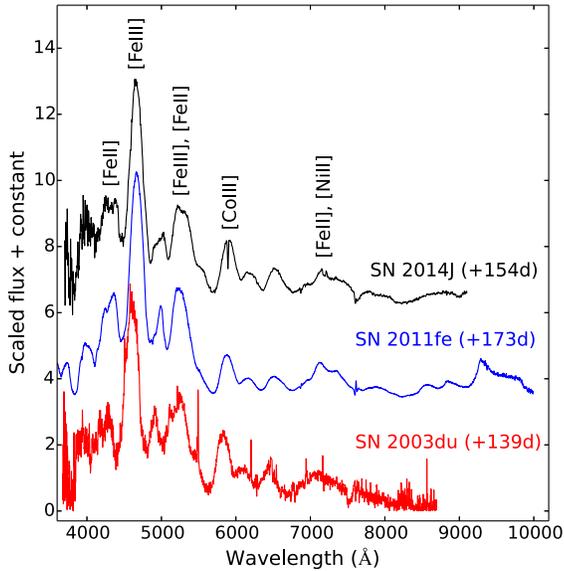}}
 \caption[]{\footnotesize Dereddened +154d nebular spectrum of SN 2014J, plotted along with spectra of SNe 2011fe and 2003du
 at similar epochs.}
 \label{fig:p150comp}
\end{figure}

\begin{figure}
\centering
\resizebox{\hsize}{!}{\includegraphics{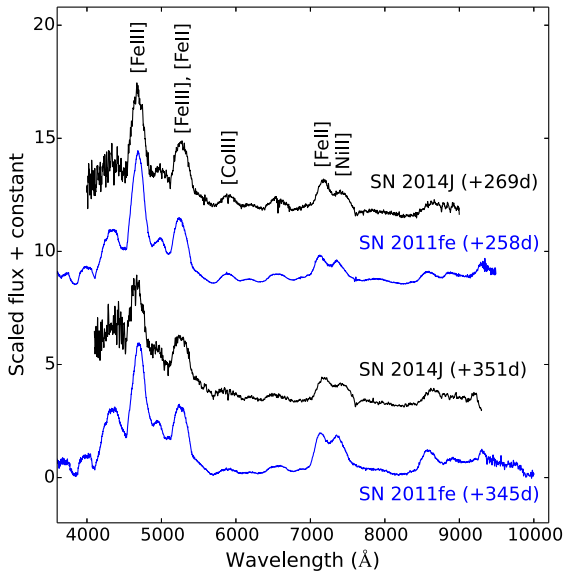}}
 \caption[]{\footnotesize Dereddened nebular spectra of SN 2014J at +269d and +351d, plotted along with nebular
 spectra of SN 2011fe at similar epochs for comparison.}
 \label{fig:nebcomp}
\end{figure}

Figure~\ref{fig:p90comp} shows the +88d spectrum of SN 2014J, along with the spectra of SNe 2011fe, 2005cf and
2003du for comparison. The spectra of all the SNe
are remarkably similar during this early nebular phase. In Figure~\ref{fig:p150comp}, 
we compare nebular spectra of SN 2014J at +154d with those of SN 2011fe \citep{Mazzali15} and
SN 2003du, whereas in Figure~\ref{fig:nebcomp}, we compare the +269d and +351d spectra of SN 2014J with those of 
SN 2011fe \citep{Mazzali15} obtained at similar epochs. 
The spectral evolution of SN 2014J is remarkably similar to SN 2011fe till $\sim 150$ days. However, their spectra show
small differences at later phases, as seen in Figure~\ref{fig:nebcomp}.
The relative strengths of the [Fe~{\sc iii}] $\lambda 4701$ feature and [Fe~{\sc ii}] $\lambda 5200$ blend in the nebular
spectra is a measure of the ionization state of the ejecta. A lower ratio of [Fe~{\sc iii}]/[Fe~{\sc ii}] indicates 
clumpiness in the ejecta \citep{Mazzali01}.
The +351d spectrum of SN 2014J shows a [Fe~{\sc iii}]/[Fe~{\sc ii}] ratio of 0.89, whereas the same ratio for the
+345d spectrum of SN 2011fe is 1.33. The relatively low [Fe~{\sc iii}]/[Fe~{\sc ii}] ratio thus 
hints towards a clumpy ejecta for SN 2014J.

We measure the velocity of the nebular emission features of SN 2014J, namely [Fe~{\sc iii}] $\lambda 4701$,
[Fe~{\sc ii}] $\lambda 7155$ and [Ni~{\sc ii}] $\lambda 7378$. The [Fe~{\sc iii}] $\lambda 4701$ feature evolves
from a velocity of $\sim -3700$ km s$^{-1}$ at +88d, 
to $\sim -1700$ km s$^{-1}$ at +351d. The [Fe~{\sc ii}] $\lambda 7155$ and [Ni~{\sc ii}] $\lambda 7378$ features
show a redshift of $\sim 1000$ and $\sim 1600$ km s$^{-1}$ respectively at +351d, unlike SN 2011fe where these
nebular features are blueshifted, as noted by \citet{Lundqvist15}.

\subsection{Spectroscopic Properties}

\begin{figure}
\centering
\resizebox{\hsize}{!}{\includegraphics{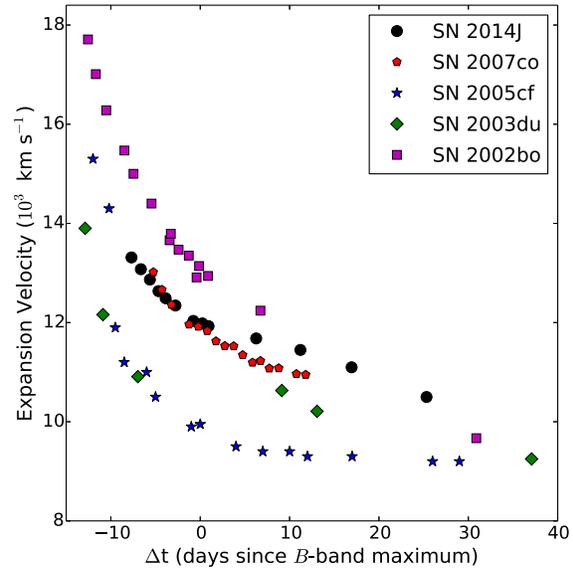}}
 \caption[]{\footnotesize Velocity evolution of Si~{\sc ii} 6355 \AA\ feature for SN 2014J, along with a few other SNe Ia
 for comparison.}
 \label{fig:velcomp}
\end{figure}

Figure~\ref{fig:velcomp} shows the velocity evolution of Si~{\sc ii} $\lambda 6355$ feature in SN 2014J during the
early phase, along with the velocity evolution of SN 2007co \citep{Blondin12}, SN 2005cf \citep{Pastorello07}, 
SN 2003du \citep{Anupama05} and SN 2002bo \citep{Benetti04}. The velocity evolution of SN 2014J in the pre-maximum 
phase is similar to the normal type Ia SN 2007co \citep{Foley11,Blondin12}.

Based on the photospheric velocity deduced from the Si~{\sc ii} $\lambda 6355$ feature near the epoch of $B$-band maximum,
\citet{Wang09a} classified SNe Ia as normal velocity (NV) or high velocity (HV) objects, where HV objects show a 
velocity $v_{phot} \gtrsim 11\,800$ km s$^{-1}$.
Fitting a spline function to the velocity
points close to maximum, we estimate $v_{phot}(B_{max}) \approx 12\,000$ km s$^{-1}$ for SN 2014J, in agreement with
\citet{Marion15}. According to the classification scheme of \citet{Wang09a}, this places SN 2014J at the border
of the HV and NV group of SNe Ia. \newline

\begin{figure}
\centering
\resizebox{\hsize}{!}{\includegraphics{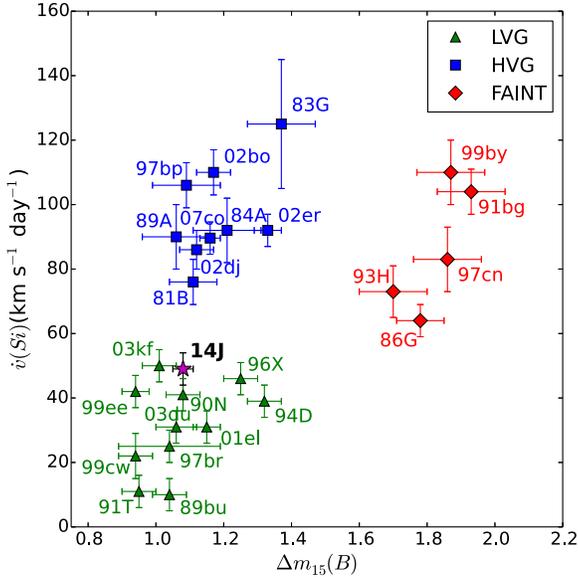}}
 \caption[]{\footnotesize Spectroscopic classification of SN 2014J according to \citet{Benetti05} scheme. The sample of SNe
 Ia have been taken from \citet{Benetti05} and \citet{Blondin12}. The position of SN 2014J is marked by the 
 filled $\bigstar$.}
 \label{fig:velgrad}
\end{figure}

\citet{Benetti05} categorized SNe Ia on the basis of the velocity gradient of the Si~{\sc ii} $\lambda 6355$
feature ($\dot{v}_{Si}$) around the epoch of $B$-band maximum. Those events showing
high velocity gradients ($\dot{v}_{Si} \gtrsim 70$ km s$^{-1}$ day$^{-1}$) are termed as HVG, whereas the ones 
showing slow velocity evolution are termed as LVG.
A third group of SNe Ia, which show low velocities but a rapid temporal velocity evolution, are categorized as FAINT.
The FAINT subclass contains sub-luminous events like SN 1991bg. The HVG comprises of normal events, while the LVG subclass
contains both normal and luminous 1991T-like events. 
Figure~\ref{fig:velgrad} shows where SN 2014J fits in with other SNe Ia according to the \citet{Benetti05} scheme.
The velocity gradient calculated for SN 2014J between +0 and +10d since $B$-band maximum, 
$\dot{v}_{Si} \approx 50$ km s$^{-1}$ day$^{-1}$.  
As pointed out by \citet{Marion15}, the measured velocity gradient places SN 2014J in the LVG group of SNe Ia, but not 
far from the boundary of HVG and LVG events. SNe 2003du and 2005cf also belong to the LVG subclass, whereas
SN 2002bo lies in the HVG group, showing high velocities and a rapid velocity evolution.
The velocity of SN 2007co matches well with SN 2014J in the pre-maximum phase, but falls off faster
post maximum. With a velocity gradient $\dot{v}_{Si} \approx 90$ km s$^{-1}$, SN 2007co lies in the HVG subclass.
The velocities of SN 2007co \citep{Blondin12} were estimated from its spectra obtained from WISeREP.

\begin{figure}
\centering
\resizebox{\hsize}{!}{\includegraphics{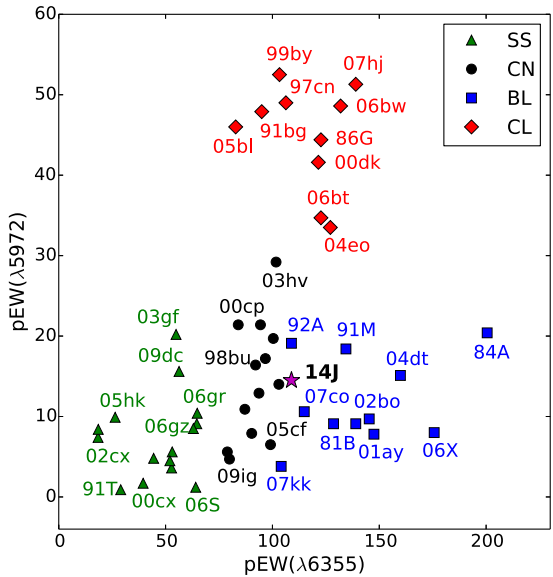}}
 \caption[]{\footnotesize Spectroscopic classification of SN 2014J according to \citet{Branch06} scheme using pEWs of 
 Si~{\sc ii} features. The sample of SNe Ia is taken from \citet{Branch06,Blondin12} and \citet{Pereira13}.
 The position of SN 2014J is marked by the filled $\bigstar$.}
 \label{fig:pewplot}
\end{figure}

\citet{Branch06} provide an alternative method of classifying SNe Ia based on their spectra. The ratio of the 
pseudo-Equivalent Widths (pEW) of Si~{\sc ii} $\lambda 5972$ and $\lambda 6355$ features in spectra near the epoch 
of $B$-band maximum forms four regions or clusters - namely core normal (CN), shallow silicon (SS), broad line (BL) 
and cool (CL). For the +0.3d spectrum of SN 2014J, we measure pEW($\lambda 5972)$ $\approx 14.5$ and 
pEW($\lambda 6355$) $\approx 108.9$.
According to the \citet{Branch06} scheme, SN 2014J lies on the border of the CN and BL subclasses of 
SNe Ia (Figure~\ref{fig:pewplot}).
SN 2007co was classified as a BL object \citep{Blondin12} and lies close to SN 2014J in the \citet{Branch06} 
scheme, as seen in Figure~\ref{fig:pewplot}.

\section{Summary and Conclusion}

We present photometric and spectroscopic observations of SN 2014J, spanning $-8$ to $+269$d and $+351$d since $B$-band
maximum, respectively. 
For a host galaxy reddening of  $E(B-V) \approx 1.24$ with $R_V \approx 1.44$ \citep{Foley14}, in addition to a
Galactic foreground reddening of $E(B-V)_{MW} \approx 0.05$, the peak $V$-band magnitude, 
$M_V(max) \approx -19.20 \pm 0.20$. The light curves show a normal decline rate of $\Delta m_{15} (B) \approx 1.08 \pm 0.03$.
Fitting a simple analytical model to the bolometric light curve provides estimates
of $M_{ej} \sim 1.3$ M$_{\odot}$, $M_{Ni} \sim 0.6$ M$_{\odot}$ and $E_k \sim 7 \times 10^{50}$ erg. 

The velocity evolution of SN 2014J, with a Si~{\sc ii} $\lambda 6355$ velocity of $\sim 12\,000$ km s$^{-1}$ at $B$-band
maximum, places it in the LVG subclass; whereas the pEW measurements of Si~{\sc ii} $\lambda 5972, \lambda 6355$
features place SN 2014J at the boundary of CN and BL subclasses of SNe Ia. 
SN 2014J shows redshifted nebular emission features of [Fe~{\sc ii}] $\lambda 7155$ ($\sim 1000$ km s$^{-1}$)
and [Ni~{\sc ii}] $\lambda 7378$ ($\sim 1600$ km s$^{-1}$) at +351d, as noted by \citet{Lundqvist15}.
The low ratio of [Fe~{\sc iii}]/[Fe~{\sc ii}] in the nebular spectra of SN 2014J indicates a lower degree of ionization
in SN 2014J, which points towards a clumpy ejecta.
 
\section*{Acknowledgements}

We thank the staff of IAO, Hanle and CREST, Hosakote, that made these observations possible. The facilities at IAO
and CREST are operated by the Indian Institute of Astrophysics, Bangalore. We also thank all HCT and AIMPOL observers
who spared part of their observing time for the observations of the supernova. 
SS is grateful to N.~K. Chakradhari for kindly providing the script used for performing template subtraction.
This work has made use of the NASA Astrophysics Data System and the NED which is operated by Jet Propulsion Laboratory, 
California Institute of Technology, under contract with the National Aeronautics and Space Administration.
We would also like to thank the anonymous referee for her/his valuable comments on the manuscript.

\bibliography{biblist}

\appendix
\section{Template Subtraction}

Pre-explosion SDSS $ugriz$ images of M82 were used in order to subtract the underlying host galaxy contribution at later
phases ($\gtrsim 60$ days since $B$-band maximum) to obtain accurate magnitudes of SN 2014J.
The SDSS images were first aligned and spatially rescaled to match the $BVRI$ images of SN 2014J.
The $ugriz$ images were then converted from counts scale to magnitude scale 
using $Image_{mag} = -2.5 \times log(Image_{counts})$.
Subsequently, the $ugriz$ images (in magnitude scale) were arithmetically transformed as prescribed by the Lupton formulae
(Lupton 2005) to obtain equivalent $BVRI$ images in the magnitude scale, which were converted back to the counts scale.
The equivalent $BVRI$ images thus obtained were used for the template subtraction. \newline

To check the accuracy of the transformations, aperture photometry was performed on the SDSS $ugriz$ images
for a small galaxy region centered around the SN location and the $ugriz$ magnitudes were calculated, which were then
transformed to $BVRI$ magnitudes using the Lupton formulae.
Next, aperture photometry was performed for the same region with the same aperture for the equivalent $BVRI$ images
obtained by performing the pixel to pixel transformation described above.
The magnitudes derived by the two methods were seen to match within 0.01 mag. 
The same test was performed on a couple of stars in the field and the magnitudes were seen to agree within 0.02 mag.

\end{document}